\documentclass[aps,twocolumn,floatfix]{revtex4}

\usepackage{graphicx}
\usepackage{dcolumn}
\usepackage{bm}
\usepackage{amssymb}

\usepackage{cleveref}

\usepackage{float}

\begin{document}


\title{RN-AdS Nontopoligical Solitons in Broken\\ Einstein-Maxwell-Higgs Theory}

\author{Ethan Honda}
\affiliation{
Vienna, VA 22180
}
\email{ehonda@alum.mit.edu}

\date{\today}

\begin{abstract}
Results are presented from numerical simulations of
the Einstein-Maxwell-Higgs equations with a broken U(1) symmetry.
Coherent nontopological soliton solutions are shown to exist that separate an
Anti-de Sitter (AdS) true vacuum interior from a  Reissner-Nordstrom (RN) false vacuum exterior.
The stability of these bubble solutions is tested by perturbing the charge of the coherent solution and evolving
the time-dependent equations of motion.
In the weak gravitational limit, the short-term stability depends on the sign 
of $(\omega/ Q) \, \partial_\omega Q$, similar to Q-balls.
The long-term end state of the perturbed solutions demonstrates a rich 
structure and is visualized using  ``phase diagrams."
Regions of both stability and instability are shown to exist for $\kappa_g \lesssim 0.015$, while
solutions with $\kappa_g \gtrsim 0.015$ were observed to be entirely unstable. 
Threshold solutions are shown to demonstrate time-scaling laws, and 
the space separating true and false vacuum end states is shown to be 
fractal in nature, similar to oscillons. 
Coherent states with superextremal charge-to-mass ratios are shown to exist and observed to 
collapse or expand, depending on the sign of the charge perturbation.
Expanding superextremal bubbles induce a phase transition to the true AdS vacuum, while
collapsing superextremal bubbles can form nonsingular strongly gravitating solutions with 
superextremal RN exteriors.

\end{abstract}


\maketitle

\section{Introduction\label{sec:Intro} }
Nontopological solitons (NTSs) are localized bound-state solutions to nonlinear field theories whose
stability is associated with a conserved Noether charge.
One of the most well-studied NTS solutions was discovered roughly thirty years ago, when
Coleman demonstrated that the flatspace Klein-Gordon equation with a nonlinear 
unbroken U(1)-symmetric potential gives rise to coherent bound states known as Q-balls \cite{Coleman1985}. 
Since that discovery, many other similar NTS solutions have been found using models with a variety
of scalar potentials (both broken and unbroken) and with the addition of gauge fields and gravity.
These soliton solutions found relevance 
in the contexts of q-stars, boson stars,  Q-ball induced solitogensis (``Q-bubbles"), baryogenesis in supersymmetric
extensions of the standard model, and other cosmological, astrophysical, and particle physics applications
%
%
\cite{SafianColemanAxenides1988,
CohenColemanGeorgiManohar1986,
JetzerBij1989,
Kusenko1997,
Kusenko1997b,
LeeStein-SchabesWatkinsWidrow1989,
GriestKolb1989,
FriemanOlintoGleiserAlcock1989,
LeviGleiser2002,
PaltiSaffinCopeland2004,
TamakiSakai2012,
GleiserThorarinson2006,
Pearce2012}. 
%

For any NTS solution to be physically relevant, it must exist long enough to interact with other
objects in the universe. As such, the stability of Q-balls and other NTS solutions
has been extensively explored both analytically and numerically.
In the original  discovery of Q-balls, Coleman used the thin-wall approximation in the case of an 
unbroken U(1) symmetry to show that coherent
solutions with energy less than the charge ($E<Q$) were stable \cite{Coleman1985}.
In the broken U(1) symmetric case, it was demonstrated that Q-balls are 
locally stable if $(\omega/ Q) \, \partial_\omega Q<0$ and locally unstable if $(\omega/ Q) \, \partial_\omega Q>0$,
where $\omega$ is the angular velocity of the phase of the complex scalar field \cite{CorreiaSchmidt2001}.
%
%
Stable Q-balls or boson stars have also been shown to exist in the presence of gravity, for 
broken and unbroken symmetries in both the thin and thick wall limits
\cite{
SakaiSasaki2007,
KawasakiKonyaTakahashi2008,
TamakiSakai2011,
Petryk2005,
Gleiser1988,
TamakiSakai2010,
TamakiSakai2011b,
SakaiTamaki2012,
TamakiSakai2014,
KleihausKunzSchneider2012}.
Stable Q-balls in the false vacuum (broken symmetry) were shown to exist 
\cite{Kusenko1997b,CorreiaSchmidt2001,SakaiSasaki2007}, 
and were dubbed ``Q-bubbles" in \cite{SakaiSasaki2007}. 
It was even shown in the broken symmetry case that gravity allows for arbitrarily small Q-balls,
where they would otherwise have been unstable in flatspace
\cite{TamakiSakai2011b}.   
An excellent review of this family of solutions, with an emphasis on strong gravitational coupling, 
is given by  \cite{LieblingPalenzuela2012}.

Oscillons and scalarons are also closely related to the solitons studied in this paper
but are not actually solitons because they are composed of a single real scalar field
and therefore do not have a conserved Noether charge.  
Oscillons created with a  double-well potential in flatspace describe 
``bubble" solutions that are of interest in the 
study of cosmological phase transitions \cite{CopelandGleiserMuller1995,HondaChoptuik2002}
and were shown  to have  fractal boundaries in the 
space of possible end states \cite{Honda2010}.
Self-gravitating scalarons discovered in \cite{NucamendiSalgado2003} were shown to
be unstable  
solutions that decayed into either an expanding vacuum bubble
or a Schwarzschild black hole.  The black hole solutions, while unstable, provide a weak 
counterexample to the no-scalar-hair conjecture
\cite{AlcubierreGonzalezSalgado2004}.  
%



While the solutions mentioned above are all obtainable from the Einstein-Maxwell-Higgs (EMH)
model in some appropriate limit (choice of potential, presence of gravity or gauge field, etc.), 
this paper discusses the existence, stability, and other properties of NTS solutions to the full
EMH theory with a broken U(1) symmetry and  Anti-de Sitter (AdS) true vacuum.
Similar $(4+1)$ models form the basis of Randall-Sundrum spacetimes, where expanding 
AdS bubbles have been shown to be candidates for brane formation \cite{Bucher2002}.

This paper begins by presenting  the general EMH formalism 
and definitions in Section \ref{sec:Formalism}. 
Section \ref{sec:Coherent} describes solutions to the coherent equations 
of motion resulting from a stationary ansatz; the basic properties 
(mass, charge, radius, central lapse) of the  solutions are presented.
%
Section \ref{sec:Evolution} describes the use of the time-dependent 
equations of motion to explore the long-term stability of the coherent solutions. 

\section{General Formalism and Definitions \label{sec:Formalism}}

The EMH action being discussed here is  given 
by
\small
\begin{equation}
S=\!\int\!\! d^4x\sqrt{|g|}\!\left( \frac{R}{16\pi \kappa_g} -\frac{F^2}{4}
  -\frac{1}{2}g^{\mu\nu}\left( D_\nu\phi\right)^*\! D_\mu\phi- V(\phi_\rho)\!\right),
\label{eqn:EMHAction}
\end{equation}
\normalsize
where 
$R$ is the Ricci curvature scalar associated with the metric $g_{\mu\nu}$,  
$F^2 = F_{\mu\nu}F^{\mu\nu}$,  
$\phi = \phi_1 + i \phi_2$ for real $\phi_1$ and $\phi_2$, and
$\phi_\rho = \sqrt{\phi_1^2 + \phi_2^2}$.  
The electromagnetic field strength tensor, the gauge covariant derivative, and the 
 $U(1)$ symmetric scalar field potential are given by
\begin{eqnarray}
F_{\mu\nu} &=& \nabla_\mu A_\nu - \nabla_\nu A_\mu, \\
D_\mu \phi &=&  \left(\nabla_\mu  - i q A_\mu\right) \phi, {\rm and} \label{eqn:GaugeCovDeriv}\\
%
%
V(\phi_\rho) &=& \sum_{n=1}^{N}\frac{\alpha_n}{2n}  \phi_\rho^{2n},
\end{eqnarray}
respectively, where $A_\mu$ is the electromagnetic vector potential with associated charge $q$. 
The action (\ref{eqn:EMHAction}) is written in terms of dimensionless variables and coordinates and
Appendix \ref{sec:AppendixB} describes how to obtain this form of the action 
from a physical dimensionful action by performing a transformation of coordinates and field variables.
Changing the dimensionless model parameter $\kappa_g$  can be interpreted 
as a rescaling of Newton's gravitational constant or the boson mass.
%
%

Varying (\ref{eqn:EMHAction}) with respect to $\phi$ gives rise to the equations of motion for 
the complex scalar field:
\vbox{
\begin{eqnarray}
\displaystyle
\nabla^\mu\nabla_\mu\phi
&=&
\displaystyle
2 i q {A}^\sigma \partial_\sigma\phi 
+ i q \phi \nabla_\sigma {A}^\sigma 
+  q^2 \phi A_\sigma A^\sigma 
\nonumber \\
&& 
+ \phi \sum_n {\alpha}_n \phi_\rho^{2n-2}.
\label{eqn:KG}
\end{eqnarray}
}
Maxwell's equations in curved spacetime are obtained by 
varying the action with respect to the vector potential and
using the antisymmetry of the field strength tensor,
%
\begin{eqnarray}
\nabla_\rho  {F}^{\sigma\rho} &=& J^\sigma 
\label{eqn:Maxwell1} \\
\partial_{\left[ \mu \right.}F_{\left. \sigma\rho \right]} &=& 0,
\label{eqn:Maxwell2} 
\end{eqnarray}
with conserved current ($\nabla_\sigma J^\sigma = 0$):
\begin{equation}
\displaystyle
J^\sigma= 
\frac{i}{2}q g^{\sigma\nu}\left( \phi^*\partial_\nu \phi - \phi \partial_\nu \phi^*\right)
+  q^2 \phi\phi^* {A}^\sigma
.
\label{eqn:Maxwell_Current}
\end{equation}

Using the standard $(3+1)$ formalism, 
the spacetime metric is given by
\begin{equation}
g_{\mu\nu}= \left(
\begin{array}{cc}
-\alpha^2 + \beta_i\beta^i &   \beta_i\\
  \beta_j &   h_{ij} \\
\end{array}
\right)
\label{eqn:ADM_metric}
\end{equation} 
for lapse function $\alpha$, shift vector $\beta^i$, and spatial metric $h_{ij}$.
Using (\ref{eqn:ADM_metric}),  
variation of (\ref{eqn:EMHAction}) with respect to $g_{\mu\nu}$ yields
a set of hyperbolic equations for the spatial metric and extrinsic curvature,
\begin{eqnarray}
h^{ik}\partial_th_{kj} &=& -2\alpha {K^i}_{j} + D^i \beta_j + D_j\beta^i \label{eqn:TimeDerivMetric}\\
\partial_t {K^i}_j &=& \pounds_\beta{K^i}_j - D^iD_j \alpha + \label{eqn:TimeDerivK} \\ 
&& \hspace{-8mm}
\alpha\left( {R^i}_j + K {K^i}_j  + 8\pi \kappa_g \left(
\frac{1}{2} {h^i}_j \left( S - \rho\right) - {S^i}_j
\right)
\right),\nonumber 
\end{eqnarray}
and the elliptic (Hamiltonian and momentum) constraint equations,
\begin{eqnarray}
^{(3)}R + K^2 - {K^i}_{j}{K^j}_{i} &=& 16 \pi\kappa_g \rho \label{eqn:Hamiltonian}\\
D_j{K^j}_{i}  - D_i K &=& 8\pi\kappa_g j_i, \label{eqn:Momentum}
\end{eqnarray}
where the energy density, momenta, and 
spatial stress tensor 
on the spatial hypersurfuce are given by 
\begin{eqnarray}
\rho &=&  n^\mu n^\nu T_{\mu\nu},  \\
j_i &=& - n^\nu T_{i \nu}, {\rm and}\\
S_{ij} &=& {h^\mu}_i {h^\nu}_j T_{\mu\nu}, 
\end{eqnarray}
in terms of the spacetime energy-momentum tensor
\begin{equation}
T_{\mu\nu} = -\frac{2}{\sqrt{-g}}\frac{\delta S_M}{\delta g^{\mu\nu}}
\end{equation}
and the normal to the spacelike hypersurface, $n^\mu = (1/\alpha,-\beta^i/\alpha)$.

In analysis of scalar field bubble dynamics, it is useful to describe a few additional quantities.
The first quantity is representative of the bubble radius and is defined as
\begin{equation}
\xi(t_j) = \left\{
\begin{array}{ll}
\textrm{max} \left( r_i\left(\phi_{TF},t_j\right) \right) & \rm{when} \ \  \phi_{TF}\in\phi(r_i,t_j)  \\
0  & \rm{otherwise},
\end{array}
\right.
\label{eqn:xidef}
\end{equation}
where $\phi_{TF} \equiv (\phi_T + \phi_F)/2$, for $\phi_T$ and $\phi_F$ being the true and
false vacuums, respectively; 
$\phi_{TF}$ satisfies  $\phi_{TF} \in \phi(r_i,t_j)$
at $t_j$ if $\phi(r_i,t_j) \leq \phi_{TF} < \phi(r_{i+1},t_j)$ for some $i$.
More simply put, 
$\xi(t)$ is the maximum radius for which the field is halfway  between 
the true and false vacuums, and zero if at time $t_j$ the field does not anywhere
equal $\phi_{TF}$.  

Another useful quantity is a normalized second moment of the magnitude of the scalar field, 
\begin{equation}
\chi(\phi_\rho,r_0) = 
\frac{ \sum_{i=0}^{N(r_0)} r_i^2 \left( (\phi_\rho)_i   - \phi_F\right)}{
 \left( \phi_T  - \phi_F\right)\sum_{i=0}^{N(r_0)} r_i^2 },
\end{equation}
where $N(r_0)$ is the index of the gridpoint corresponding to $r_0$.  
$\chi(\phi_\rho,r_0)$ gives a measure of volume of space within $r\leq r_0$ that 
is occupied by the true or false vacuum,
normalized to give zero for the false vacuum and
one for the true vacuum.

Finally, the choices for $\alpha_n$ are such that $V(\phi)$ is a broken U(1) symmetric potential, 
$\alpha_1 = 1$, $\alpha_2 = -5/2$, and $\alpha_3 = 1$, with 
false vacuum at $\phi_F=0$ with $V(\phi_F) = 0$, and with
AdS true vacuum at  $\phi_T = \sqrt{2}$ with $V(\phi_T) = -1/6$.
Unless otherwise explicitly stated, the value of the charge is taken to be $q=0.1$.

%
%
%
%

\section{Coherent Reissner-Nordstrom Anti-de Sitter NTS Solutions  \label{sec:Coherent}}
This section discusses the stationary spherically symmetric coherent solutions to  
(\ref{eqn:KG}),
(\ref{eqn:Maxwell1}),
(\ref{eqn:Maxwell2}),
(\ref{eqn:TimeDerivMetric}),
(\ref{eqn:TimeDerivK}), 
(\ref{eqn:Hamiltonian}), 
and 
(\ref{eqn:Momentum}). 
The polar-areal slicing conditions ($b=1$, ${\rm Tr}K = {K^r}_r$) fix the spacetime gauge 
and simplify the metric to
\begin{equation}
ds^2 = -\alpha^2(r) dt^2 + a^2(r) dr^2 + r^2 d\Omega,
\end{equation}
where $d\Omega = d\theta^2 + \sin^2\theta d\phi^2$.
The Maxwell gauge freedom is set using the Lorentz gauge condition,
\begin{equation}
\nabla_\mu A^\mu = 0,
\end{equation} 
and the electric field is implicitly defined such that 
\begin{equation}
F_{tr} = -\alpha a E_r.
\end{equation}
Assuming the scalar field is described by the coherent ansatz,
\begin{equation}
\phi(t,r) = \phi_\rho(r)  e^{i \omega t},
\end{equation}
the coherent equations of motion are given by
\begin{eqnarray}
\frac{1}{r^2}\partial_r\left( \frac{\alpha r^2\Phi_\rho}{a}  \right) &=& 
-\frac{a}{\alpha} \phi_\rho u^2 
+ \alpha a \sum_n \alpha_n\phi_\rho^{2n-1}, \label{eqn:CoherentBigPhi}\\
\partial_r\phi_\rho &=& \Phi_\rho, \label{eqn:CoherentPhi}\\
\frac{1}{r^2}\partial_r \left( r^2 E_r\right) &=&  J^t,  \label{eqn:CoherentEr}\\
\partial_r u &=&   \alpha a q E_r, \label{eqn:CoherentU}\\
\frac{a'}{a} &=& \frac{1- a^2}{2r} + 4\pi \kappa_g r a^2 \rho,  \label{eqn:CoherentA}\\
\frac{\alpha'}{\alpha} &=& \frac{a^2 - 1}{2r} + 4\pi\kappa_g r a^2 {S^r}_r, \label{eqn:CoherentAlpha}
\end{eqnarray}
where
\begin{eqnarray}
u &=& q A_t - \omega,\\
 J^t &=&  \frac{q u a \phi_\rho^2 }{\alpha}, \label{eqn:Jt}\\
\rho 		&=&   \frac{E_r^2}{2}  + \frac{\Phi_\rho^2}{2a^2} + 
 \frac{u^2  \phi_\rho^2}{2 \alpha^2}+ V, \\
{S^r}_r	&=&  -\frac{E_r^2}{2}  + \frac{\Phi_\rho^2}{2a^2} + 
 \frac{u^2  \phi_\rho^2}{2 \alpha^2} - V, 
\end{eqnarray}
and the conserved charge and mass are given by
\begin{eqnarray}
Q 	&=& 4\pi\int_0^{r_b} dr r^2 J^t \\
M  	&=& 4\pi \int_0^{r_b} dr r^2 \rho, 
\end{eqnarray}
where $r_b$ is the radial coordinate value of the outer boundary of the computational domain.
While $M$ is usually a decent approximation for the total mass, for precise
measurements using the total mass, one should use the mass at spatial infinity, 
\begin{equation}
M_\infty(r) = 
\left( \frac{r}{2\kappa_g}\right)
\left( 1 - a^{-2} + \frac{\kappa_g Q^2}{4\pi r^2}\right),
\end{equation}
which is obtained by matching the spacetime to the Reissner-Nordstrom (RN) geometry. 
Since these solutions are bound states of finite extent with radii approximated by $\xi$, 
$M_\infty(r)$ tends to approach a constant very rapidly for $r>\xi$.  When not explicitly stated, 
$M_\infty$ is evaluated at $r=r_b$.

It is helpful to note that the equations of motion are invariant under 
the multiplicative rescaling of $\alpha$ and $u$,
\begin{eqnarray}
(u,\alpha) & \rightarrow& (k_1 \alpha,k_1 u), \label{eqn:Rescale1}
\end{eqnarray}
and the additive shifting of $A_t$ and $\omega$,
\begin{eqnarray}
(A_t,\omega) &\rightarrow& (A_t + k_2 , \omega + qk_2), \label{eqn:Shift1}
\end{eqnarray}
for real constants $k_1$ and $k_2$.  

\begin{figure}[t]
\begin{center}
\hbox{
\includegraphics[width=8cm,height=8cm]{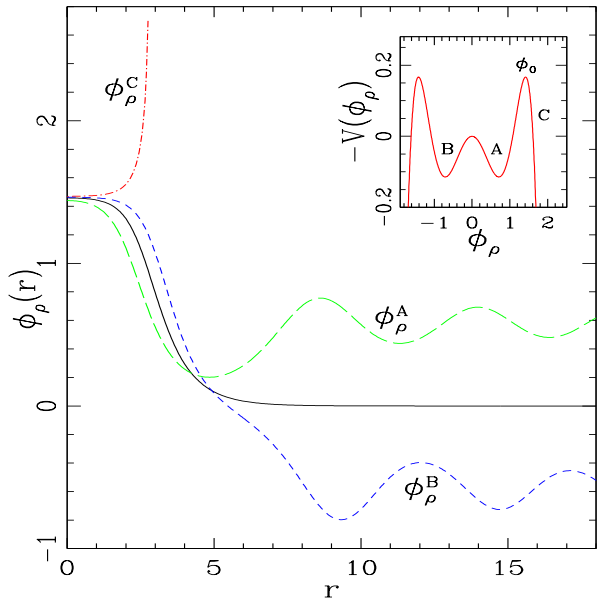}
}
\end{center}
\vspace{-10mm}
\caption{  
Solutions to equations 
(\ref{eqn:CoherentBigPhi}),
(\ref{eqn:CoherentPhi}),
(\ref{eqn:CoherentEr}),
(\ref{eqn:CoherentU}), 
(\ref{eqn:CoherentA}), 
and
(\ref{eqn:CoherentAlpha}) with different $\phi_0$ values.
Solutions $\phi^A_\rho$ and $\phi^B_\rho$ correspond, respectively, to  ``under"-shot and ``over"-shot 
solutions that oscillate in the minima of $-V(\phi_\rho)$.    
Solution $\phi^C_\rho$ is an example of a runaway solution that ``blows up" to infinity.
 $\phi_\rho(r)$ is the solution corresponding to the coherent bound state.
Although the behavior observed is generic, these solutions are for $q = 0.05$,
$\kappa_g = 0.001$, and $u_0=-0.5$.  The inset shows $-V(\phi_\rho)$.
} 
\label{fig:ShootExit}
\end{figure}

The coherent equations of motion are solved using a second-order finite difference 
code utilizing standard double precision \cite{IEEE754-2008} variables.
Solutions are obtained by setting $u_0 \equiv u(r=0)$ and using 
$\phi_0\equiv \phi(r=0)$ as a shooting parameter.  Figure \ref{fig:ShootExit}  
shows the different possible outcomes for shooting solutions, depending on
the trial values for $\phi_0$.  
To obtain a coherent solution, $\phi(r)$, one starts with 
two values of $\phi_0$ that yield solutions like $\phi^A_\rho$ and $\phi^B_\rho$ that 
oscillate in different local minima of $-V(\phi_\rho)$ .
Bisecting between two such solutions yields the eigenvalue solution $\phi(r)$ 
that asymptotically approaches the false vacuum.
This also results in $u(r)$ and $\alpha(r)$ solutions that approach 
constant values for large $r$. 
%
$\alpha(r)$ and $u(r)$ are then rescaled using (\ref{eqn:Rescale1}) 
to set $\alpha = 1/a$ at the outer boundary. 
Finally, $\omega$ and $A_t(r)$ are then determined from $u(r)$ using 
(\ref{eqn:Shift1}) such that $A_t(\infty) = 0$.
These charged scalar field bound-states that 
interpolate between vacua of a U(1) symmetric potential
are similar to the ``Q-bubble" solutions discussed  in 
\cite{Kusenko1997b,
CorreiaSchmidt2001,
SakaiSasaki2007}.
Since these bubble solutions are nontopological solitons with an AdS interior and a 
RN exterior, they are referred to here as Reissner-Nordstrom Anti-de Sitter
nontopological solitons (RN-AdS NTSs).

\begin{figure}[t]
\begin{center}
\hbox{
\includegraphics[width=8cm,height=12cm]{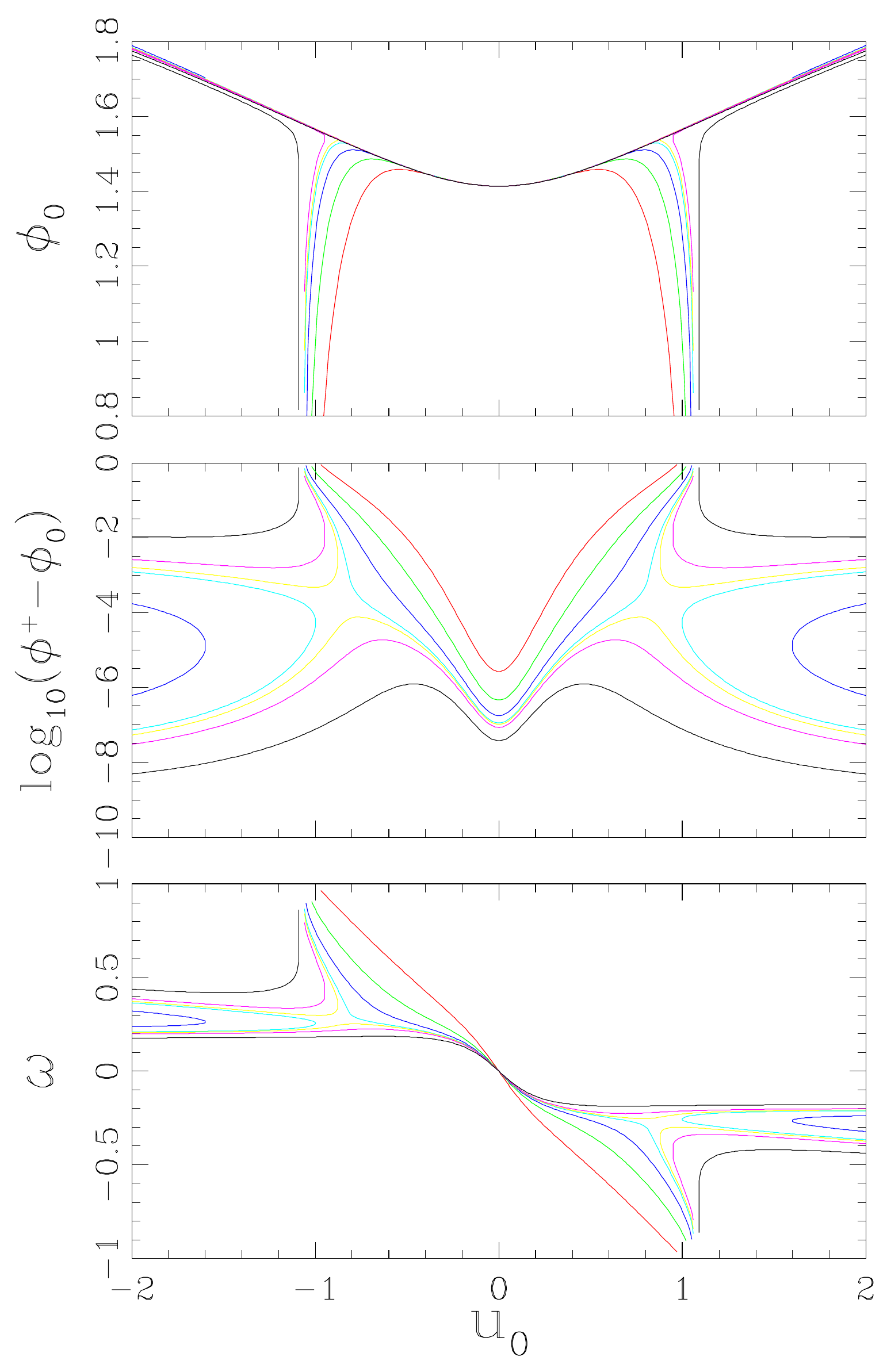}
}
\end{center}
\vspace{-10mm}
\caption{  
Plots of $\phi_0$, $\log_{10}(\phi^+  - \phi_0 )$, and $\omega$
as a function of $u_0$
for coherent bound-state solutions with 
$\kappa_g = \{ 0.001,  0.008, 0.011, 0.01225, 0.0125, 0.013, 0.015\}$
plotted in red, green, blue, cyan, yellow, magenta, and black, respectively.
} 
\label{fig:Psi_vs_U0}
\end{figure}

The parameter space of coherent solutions is explored by varying $u_0$ 
for various gravitational couplings, $\kappa_g$. 
Figure \ref{fig:Psi_vs_U0} (top) shows the values of $\phi_0$ that result in 
coherent bound states. 
Since there are 
many solutions close to one another in $\phi_0$-space, $\log_{10}(\phi^+-\phi_0)$
is also plotted (Figure \ref{fig:Psi_vs_U0}, middle), 
where $\phi^+$ is the lowest value of $\phi_0$ above which all solutions ``blow up" 
like the $\phi^C_\rho$ solution in Figure \ref{fig:ShootExit}.
Figure \ref{fig:Psi_vs_U0} (bottom) shows the values of $\omega$ for the different coherent 
solutions.
As is common with Q-balls \cite{TamakiSakai2010}, 
a ``catastrophe" clearly can be seen between the
$\kappa_g = 0.1225$ and  $\kappa_g = 0.125$ solutions, around $u_0\approx\pm 0.9$,
where the topology of the 
solutions in $u_0$-space can be seen to change.

\begin{figure}[t]
\begin{center}
\hbox{
\includegraphics[width=8cm,height=12cm]{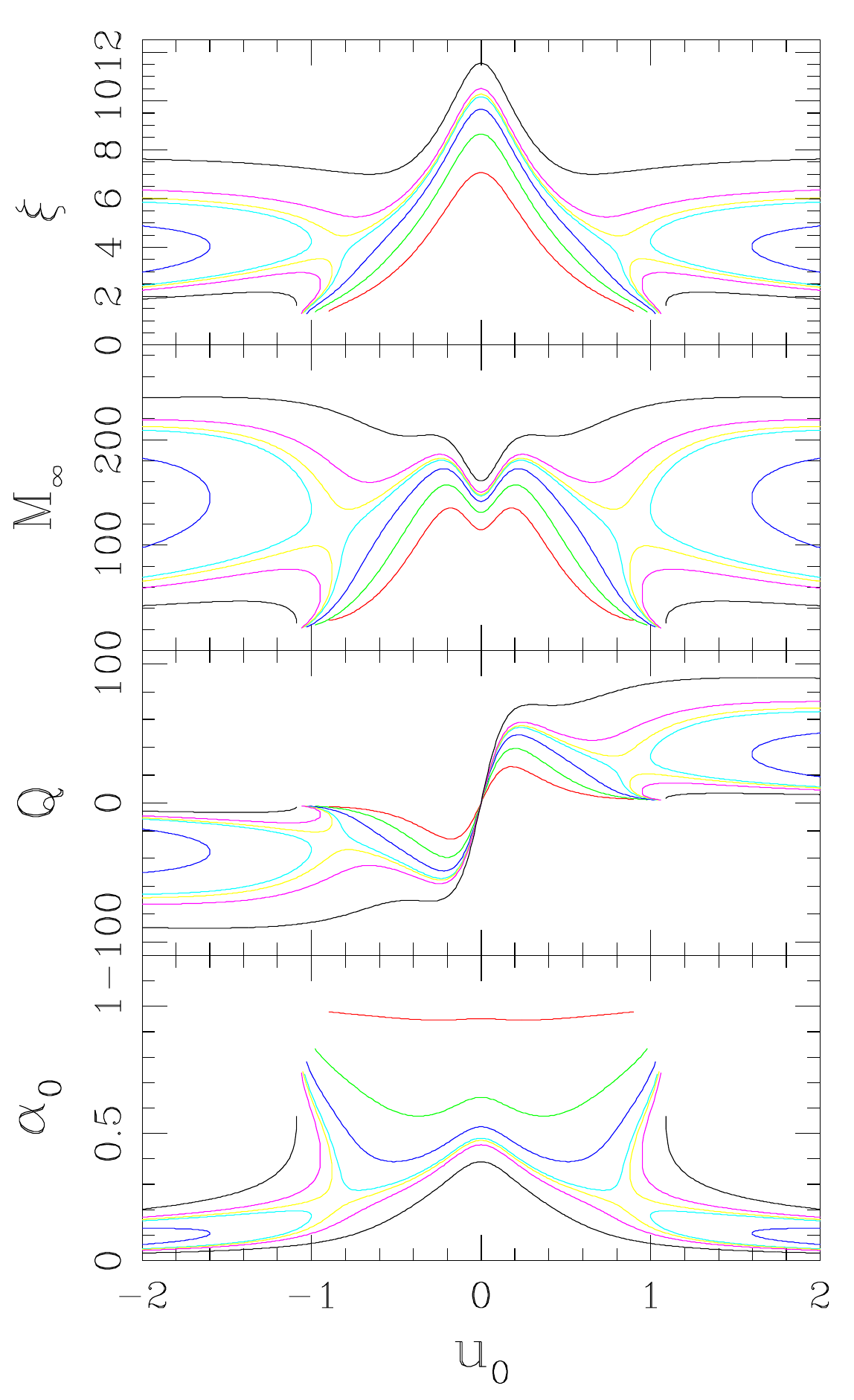}
}
\end{center}
\vspace{-10mm}
\caption{  
Plots of radius ($\xi$), mass ($M_\infty$), charge ($Q$), and 
central lapse ($\alpha_0$) as a function of $u_0$ for 
coherent bound-state solutions for 
$\kappa_g = \{ 0.001,  0.008, 0.011, 0.01225, 0.0125, 0.013, 0.015\}$
plotted in red, green, blue, cyan, yellow, magenta, and black, respectively.
} 
\label{fig:RMQL_vs_U0}
\end{figure}

\begin{figure}[t]
\begin{center}
\hbox{
\includegraphics[width=8cm,height=12cm]{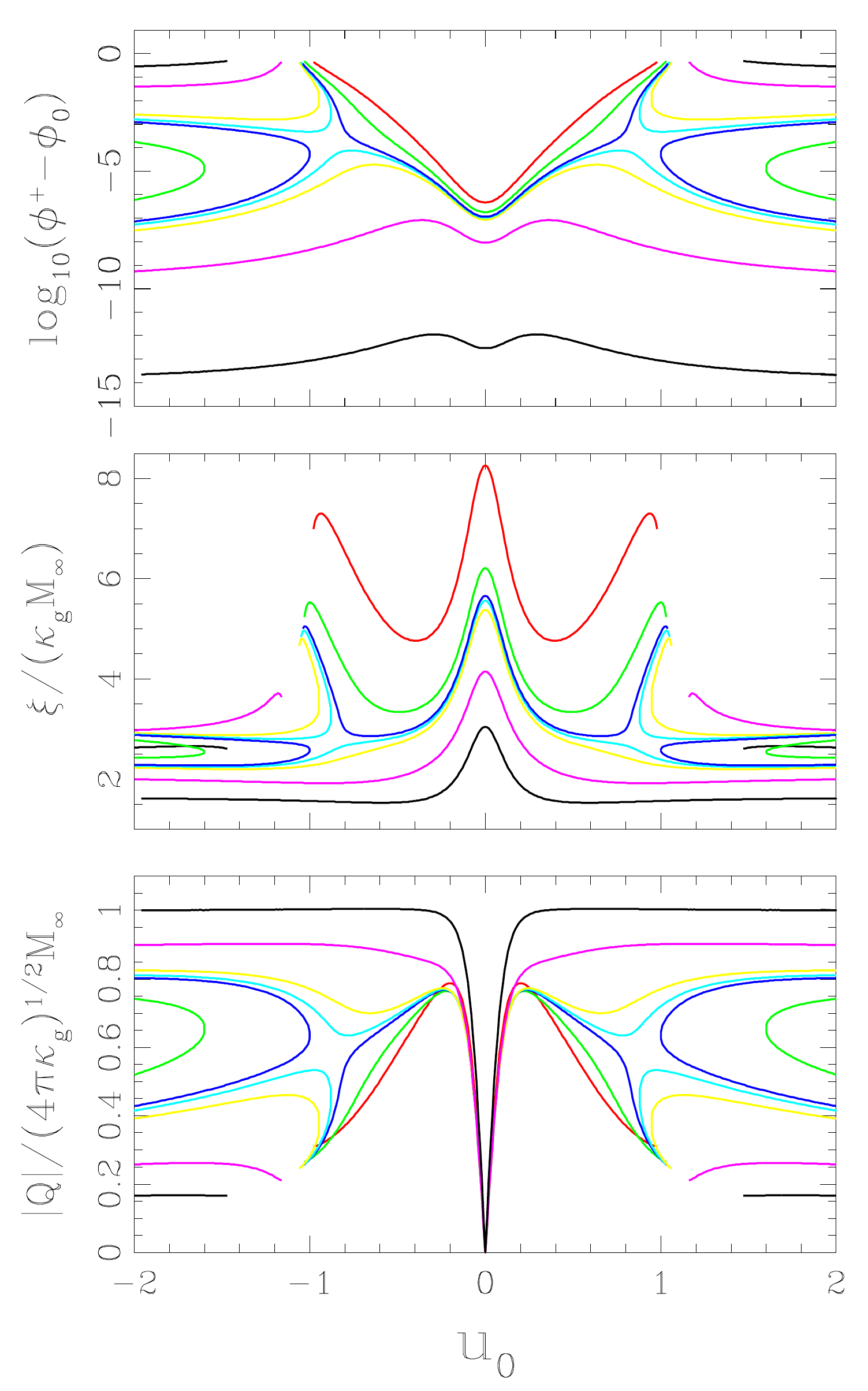}
}
\end{center}
\vspace{-10mm}
\caption{ 
Plots of $\log_{10}(\phi^+  - \phi_0 )$, $\xi/(\kappa_g M_\infty)$,
and $|Q|/(4\pi\kappa_g)^{1/2} M_\infty$ as a function of $u_0$ for 
coherent bound-state solutions for 
$\kappa_g = \{ 0.008, 0.011, 0.01225, 0.0125, 0.013, 0.018,0.029\}$
plotted in red, green, blue, cyan, yellow, magenta, and black, respectively.
} 
\label{fig:Schwarz_vs_U0}
\end{figure}

\begin{figure}[t]
\begin{center}
\hbox{
\includegraphics[width=8cm,height=8cm]{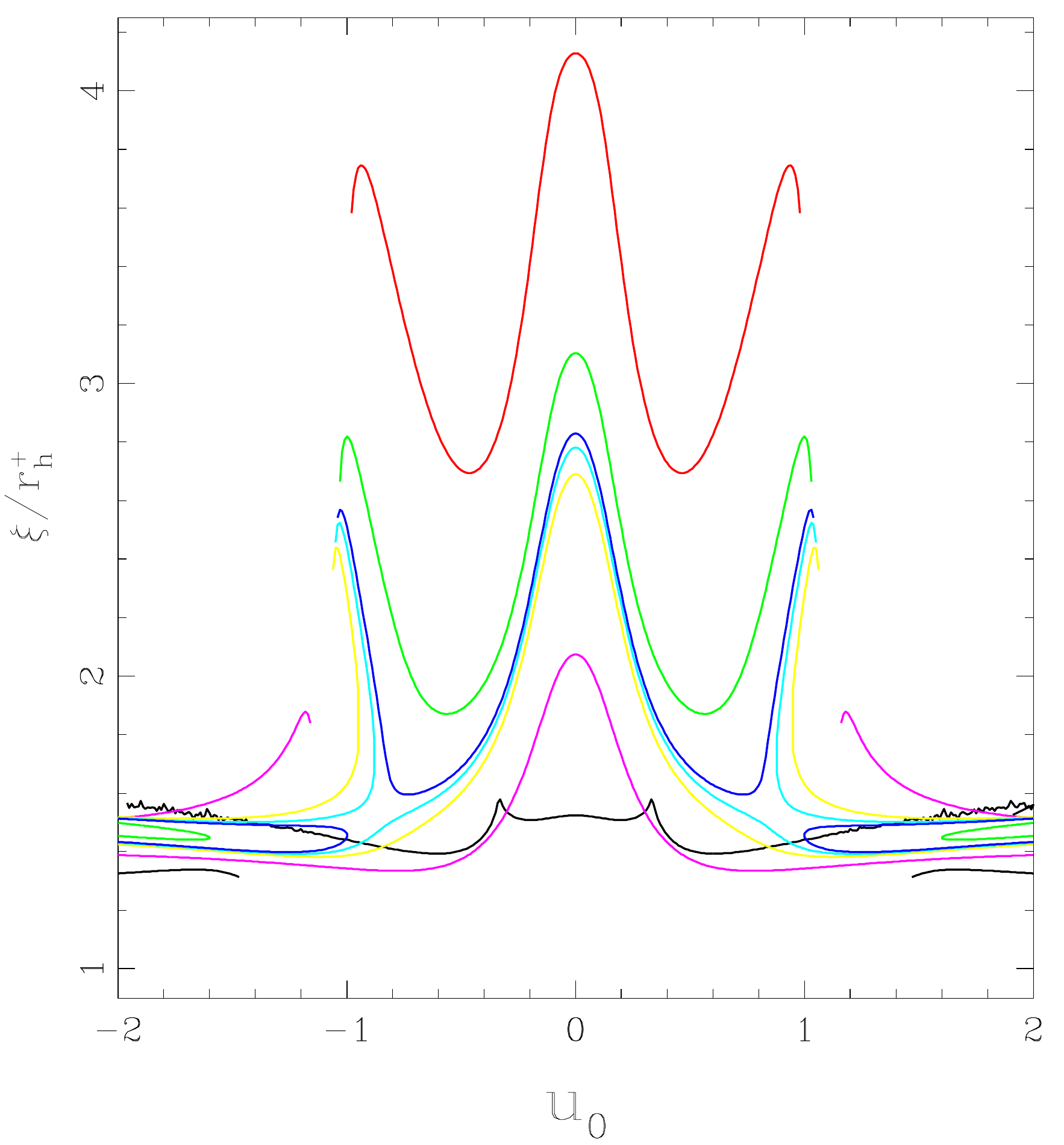}
}
\end{center}
\vspace{-10mm}
\caption{ 
Plots of  $\xi/r_h^+$  as a function of $u_0$ for 
coherent bound-state solutions for 
$\kappa_g = \{ 0.008, 0.011, 0.01225, 0.0125, 0.013, 0.018,0.029\}$
plotted in red, green, blue, cyan, yellow, magenta, and black, respectively.
For the $\kappa_g = 0.029$ solutions, the evidence of loss of numerical precision from 
using the double-precision code can be seen.
} 
\label{fig:rplus_vs_U0}
\end{figure}

%
Figure \ref{fig:RMQL_vs_U0} shows the values of 
the radius ($\xi$), mass ($M$), charge ($Q$), and central lapse ($\alpha_0$), for a range of $u_0$
and gravitational couplings $\kappa_g$.
Due to a quadratic dependence on $u$, the values of $\xi$, mass, and $\alpha_0$ 
are seen to be symmetric in $u_0$, while linear dependence 
of the charge density (\ref{eqn:Jt}) on $u$ yields a total charge that is antisymmetric in $u_0$.
Figures \ref{fig:Schwarz_vs_U0} and \ref{fig:rplus_vs_U0}
show that for gravitational couplings $\kappa_g \gtrsim 0.011$, the
NTS solutions are gravitationally strong-field  solutions with radii on the order of their 
RN horizon radii, where the inner (--) and outer (+) RN horizons are given by
\begin{eqnarray}
r_h^\pm &=& \kappa_gM_\infty \pm \sqrt{\kappa_g^2M_\infty^2 - \frac{\kappa_gQ^2}{4\pi}}  \\
&=& \kappa_gM_\infty \left( 1 \pm \sqrt{1-\Xi^2} \right)
\end{eqnarray}
and where 
\begin{equation}
\Xi = \frac{Q}{(4\pi\kappa_g)^{1/2} M_\infty}
\end{equation}
is the charge-to-mass ratio. $\Xi$ is defined such that  
for subextremal solutions, $|\Xi|<1$ and $r_h^+$ is between $\kappa_gM_\infty$ and $2\kappa_gM_\infty$,
for extremal solutions, $\Xi=\pm 1$ and $r_h^-$ and $r_h^+$ are coincident at  $\kappa_gM_\infty$, 
and for superextremal solutions, $|\Xi|>1$ and there are no real $r_h^\pm$.
While one may question the use of the term ``extremal" for nonsingular NTS solutions 
without horizons ($\xi > r_h^+$), 
coherent NTS solutions with $|\Xi| = 1$ are extremal in that they are non-interacting with other
like-charged extremal solutions because
their gravitational attraction is balanced by their Coulombic repulsion.  
Additional evidence of the strong-field nature of these solutions
($0.01\lesssim \kappa_g \lesssim 0.03$) 
is that the central lapse values are typically $0.1 \lesssim \alpha_0 \lesssim 0.5$, 
indicating significant gravitational time dilation relative to observers at spatial infinity where
$\alpha=1$.

\begin{figure}[t]
\begin{center}
\hbox{
\includegraphics[width=8cm,height=12cm]{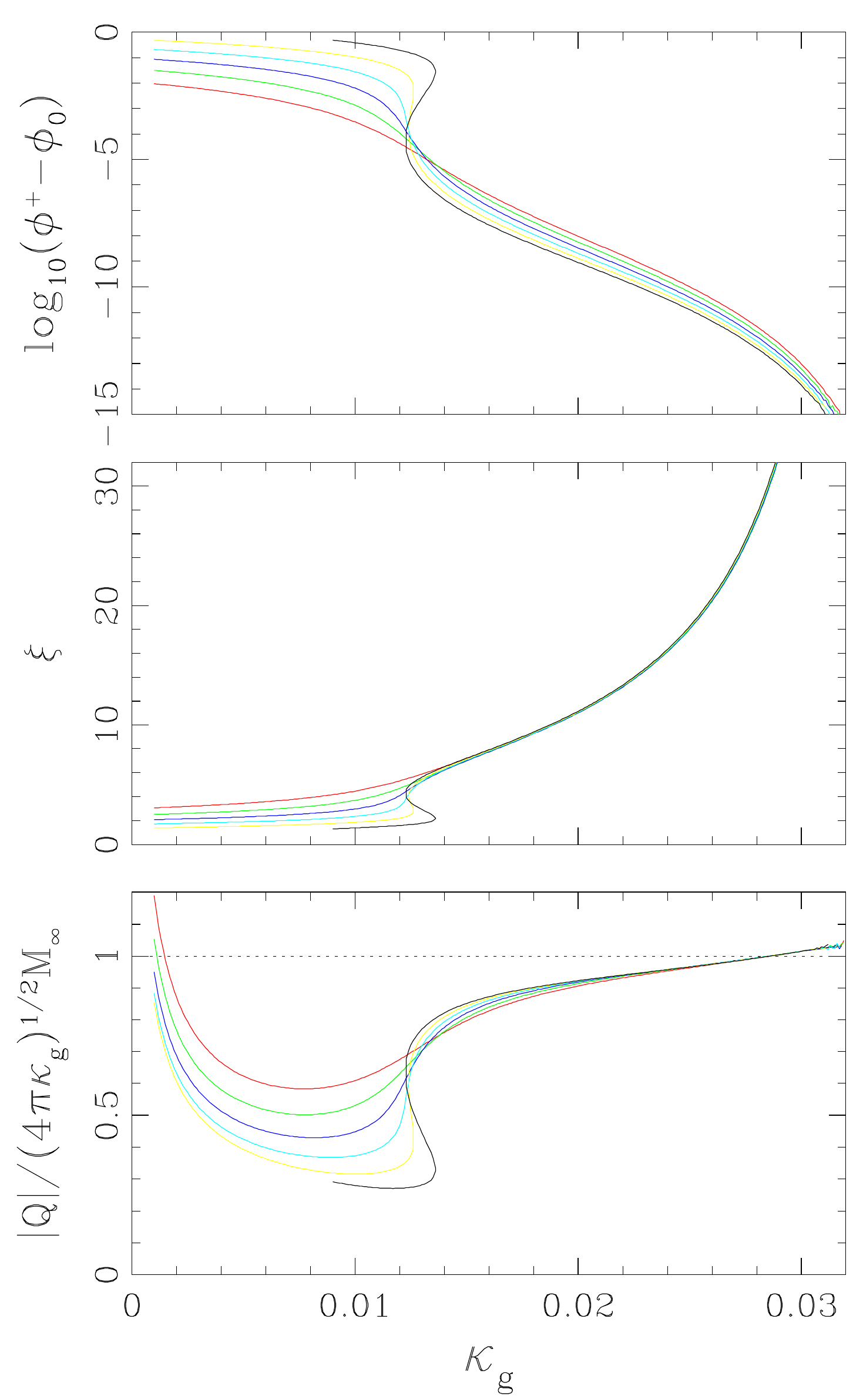}
}
\end{center}
\vspace{-10mm}
\caption{  
Plots of  \hbox{$\log (\phi^+_0-\phi_0 )$}, radius ($\xi$), and $|Q|/(4\pi\kappa_g)^{1/2} M_\infty$ 
as a function of gravitational coupling
$\kappa_g$ for \hbox{$u_0=\{0.5, 0.6, 0.7, 0.8, 0.9, 1.0 \}$}, plotted in red, green, blue, cyan,
yellow, and black, respectively. 
The dashed line in the bottom graph denotes the extremality condition $|Q|/(4\pi\kappa_g)^{1/2} M_\infty = 1$.
For values of $\kappa_g \gtrsim 0.03$ loss of numerical precision from using the double prevision code 
can be seen.
} 
\label{fig:RQovMLogPsi_vs_kg}
\end{figure}
%
Solutions with $|\Xi| \geq 1$ are of particular interest,
given their potential to form extremal black holes or naked singularities.
Figure \ref{fig:RQovMLogPsi_vs_kg} shows that coherent solutions with $|\Xi| \geq 1$ appear to exist
for values of $\kappa_g \lesssim 0.002$ and $\kappa_g \gtrsim 0.028$.
%
Solutions with $\kappa_g\lesssim 0.002$ are not likely to couple to gravity strongly enough to 
result in collapse to within $r_h^+$;
the $\kappa_g\gtrsim 0.028$ solutions are promising candidates 
for superextremal collapse but appear to show a loss of numerical precision 
when using standard double-precision variables.
%

\begin{figure}[t]
\begin{center}
\hbox{
\includegraphics[width=8cm,height=10cm]{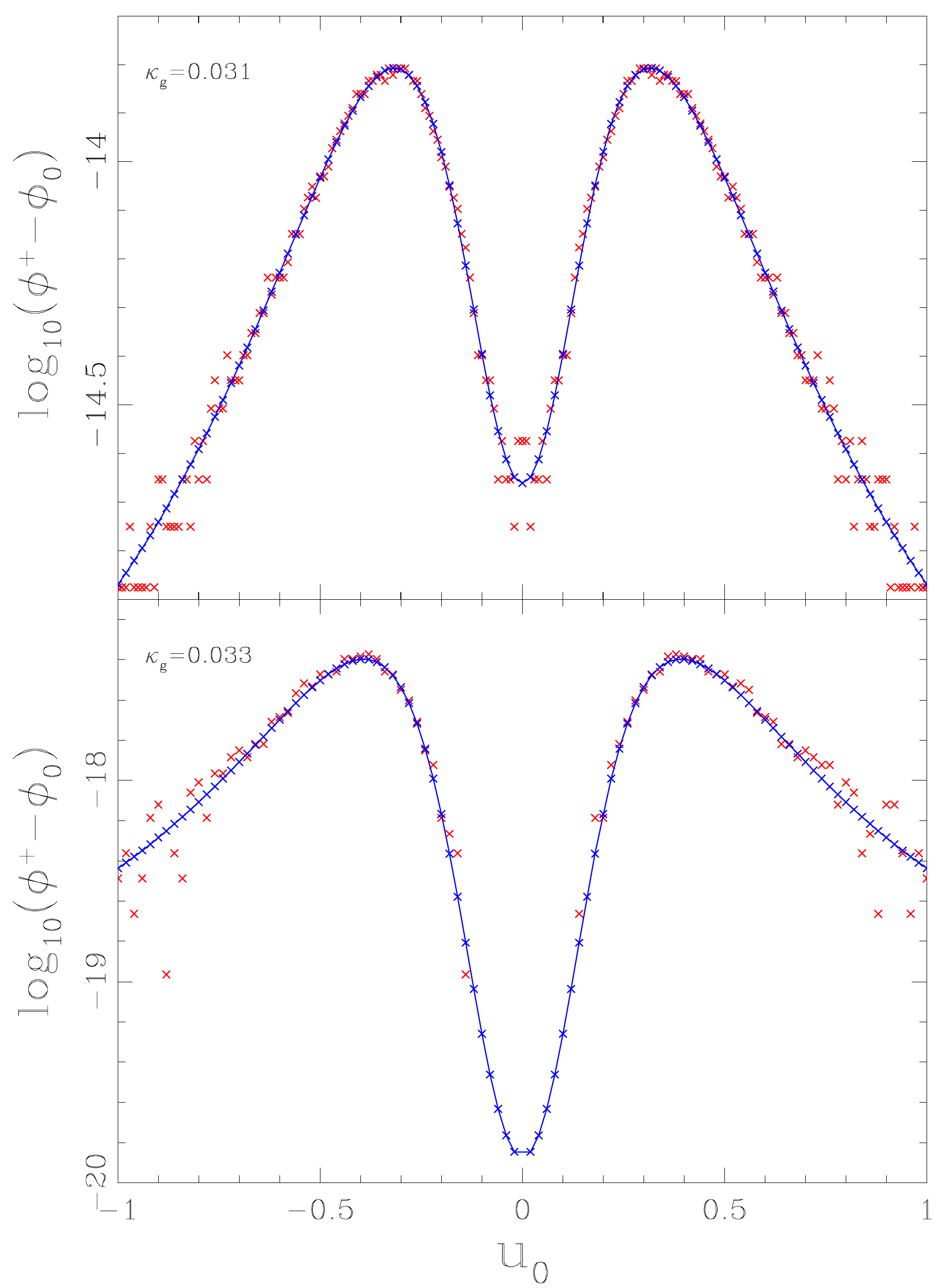}
}
\end{center}
\vspace{-10mm}
\caption{  
Plots of \hbox{$\log (\phi^+_0-\phi_0 )$} as a function of $u_0$ for two different values 
of gravitational coupling with simulations of different numerical precision.
The top graph is for $\kappa_g=0.031$, where the red x's are using standard 53-bit double precision 
variables and the blue connected x's are using 64-bit precision variables.
The bottom graph is for $\kappa_g=0.033$, where the red x's are using 64-bit precision 
variables and the blue connected x's are using 96-bit precision variables. 
} 
\label{fig:PrecisionEffects}
\end{figure}
%
Standard double-precision variables are encoded with 64 bits and have 53 bits of
precision dedicated to the mantissa of the real number they are representing. Such variables cannot distinguish 
between two different real numbers to more than one part in $10^{53\log_{10}\! 2} \approx 10^{16}$.
This effect begins to become apparent when $\log_{10}\left( \phi^+ - \phi_0\right) \lesssim -15$
and the ability for a double-precision code to resolve additional large-$\kappa_g$ coherent solutions is lost.  
For additional precision, a code was created that uses $n\times 32$ bits
of precision.
Figure \ref{fig:PrecisionEffects} compares the use of 64 bits of precision to the
standard 53 bits, and the use of 96 bits of precision to 64 bits.  
The 64-bit precision code can fine-tune solutions to one part in $10^{64\log_{10}\! 2} \approx 10^{19}$,
and the 96-bit precision code can fine-tune to one part in $10^{96\log_{10}\! 2} \approx 10^{29}$.

\begin{figure}[t]
\begin{center}
\hbox{
\includegraphics[width=8cm,height=12cm]{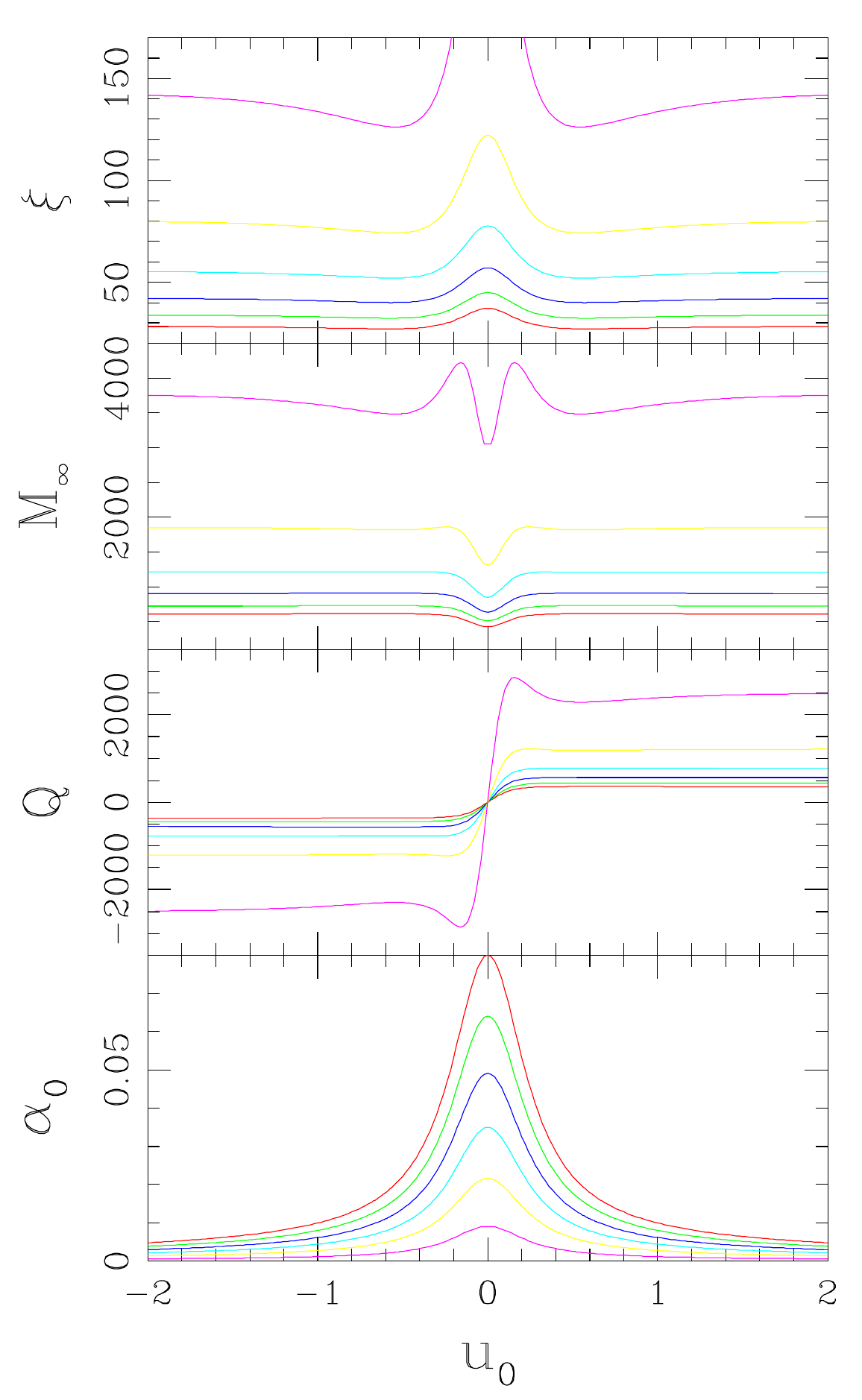}
}
\end{center}
\vspace{-10mm}
\caption{  
Plots of radius ($\xi$), mass ($M_\infty$), charge ($Q$), and 
central lapse ($\alpha_0$) as a function of $u_0$ for 
coherent bound-state solutions for 
$\kappa_g = \{ 0.028, 0.029, 0.030, 0.031, 0.032, 0.033\}$
plotted in red, green, blue, cyan, yellow, and magenta, respectively.
} 
\label{fig:RMQL_vs_U0_HighKappa}
\end{figure}

\begin{figure}[t]
\begin{center}
\hbox{
\includegraphics[width=8cm,height=12cm]{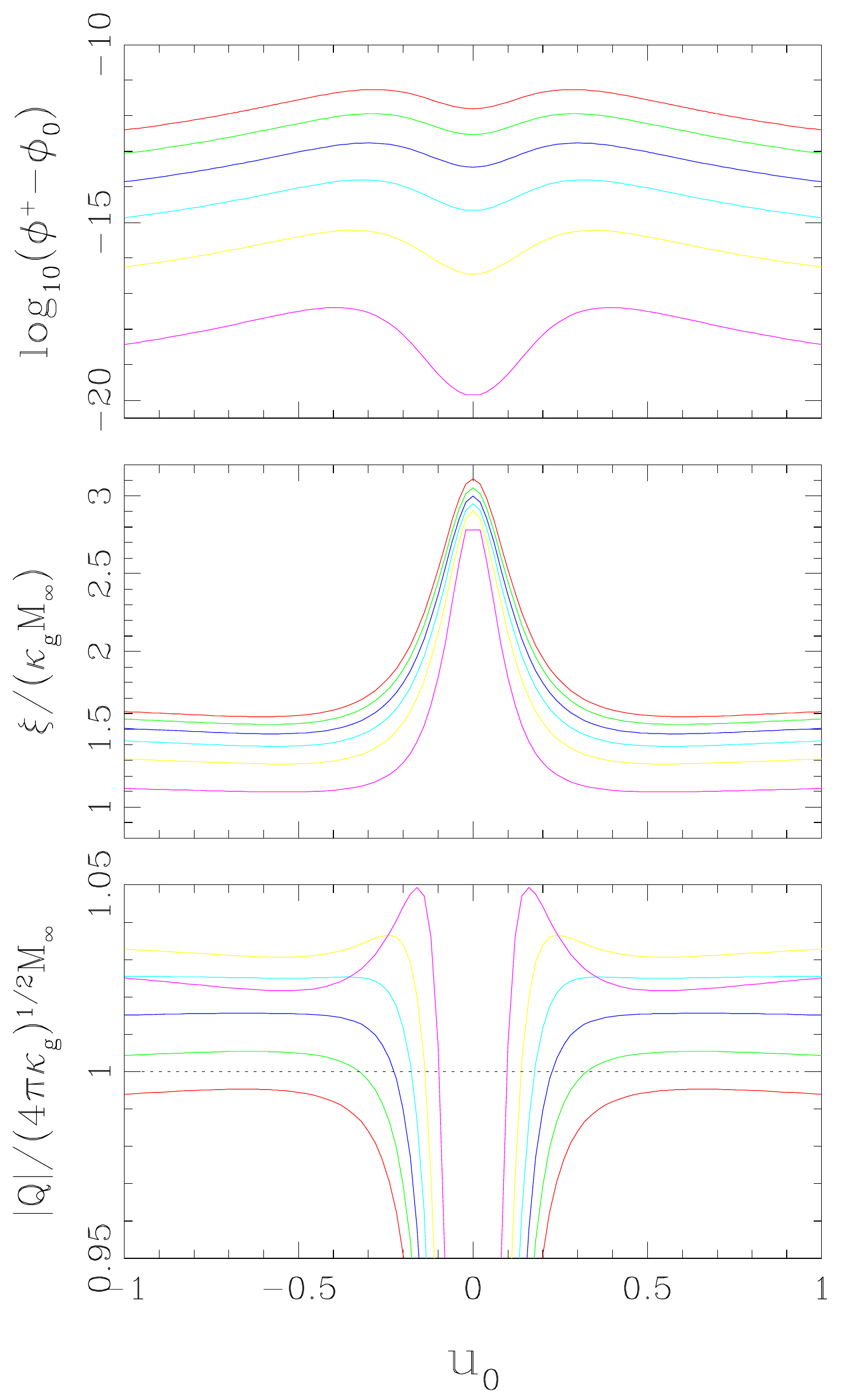}
}
\end{center}
\vspace{-10mm}
\caption{ 
Plots of  \hbox{$\log (\phi^+_0-\phi_0 )$}, $\xi/(\kappa_g M_\infty)$, 
and $|Q|/(4\pi\kappa_g)^{1/2} M_\infty$ as a function of $u_0$ for 
coherent bound-state solutions for 
$\kappa_g = \{ 0.028, 0.029, 0.030, 0.031, 0.032, 0.033\}$
plotted in red, green, blue, cyan, yellow, and magenta, respectively.
The dashed line in the bottom graph denotes the extremality condition $|Q|/(4\pi\kappa_g)^{1/2} M_\infty = 1$
and separates sub- and superextremal solutions.
} 
\label{fig:Schwarz_vs_U0_HighKappa}
\end{figure}

\begin{figure}[t]
\begin{center}
\hbox{
\includegraphics[width=8cm,height=12cm]{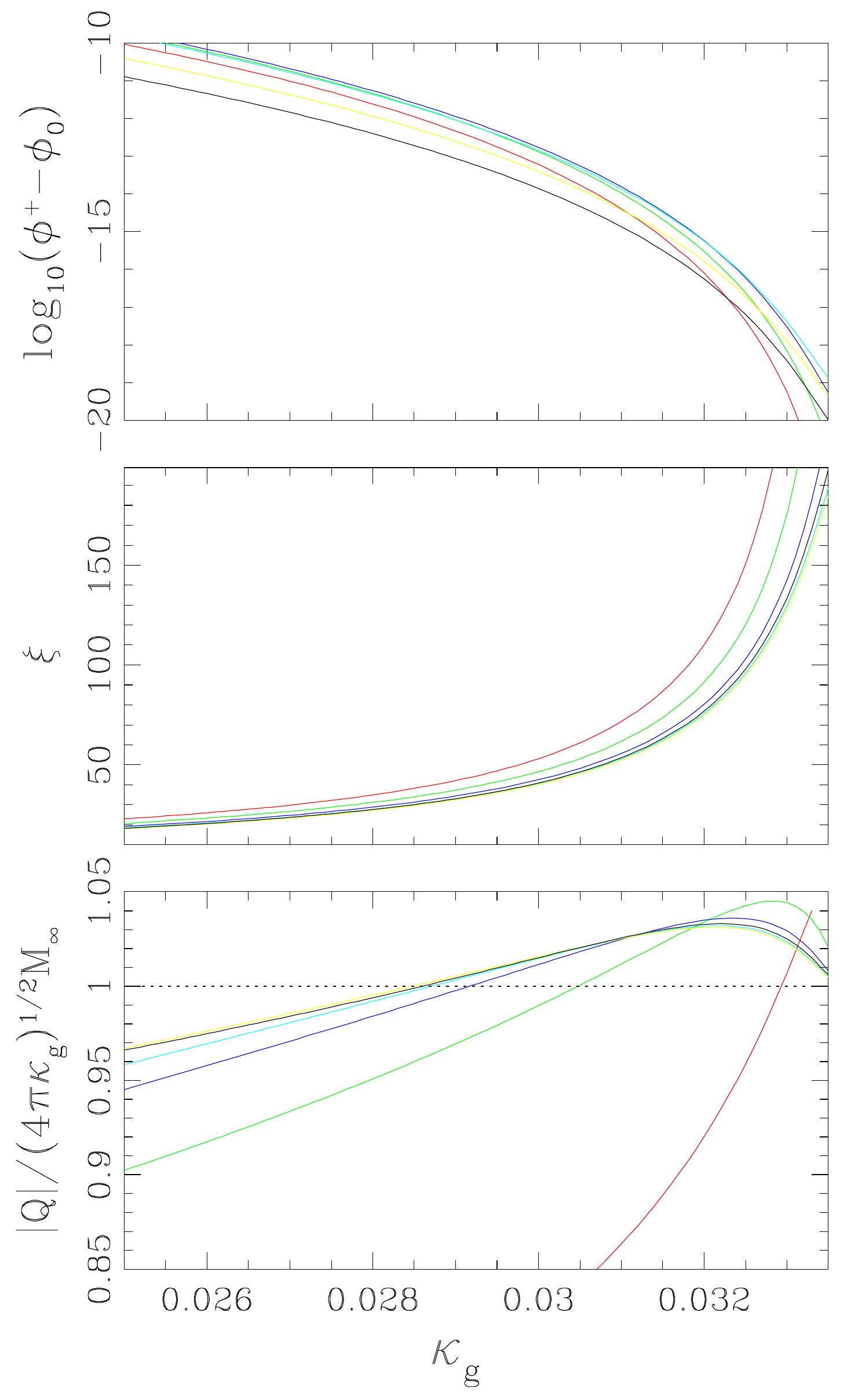}
}
\end{center}
\vspace{-10mm}
\caption{  
Plots of $\xi$, \hbox{$\log_{10}(\phi^+_0 - \phi_0 )$}, and $|Q|/(4\pi\kappa_g)^{1/2} M_\infty$ 
as a function of $\kappa_g$ for coherent bound-state solutions for 
\hbox{$u_0=\{0.1, 0.2, 0.3, 0.4, 0.7, 1.0 \}$} 
in red, green, blue, cyan, yellow, and black, respectively.  
The dashed line in the bottom graph denotes the extremality condition $|Q|/(4\pi\kappa_g)^{1/2} M_\infty = 1$
and separates sub- and superextremal solutions.
} 
\label{fig:RQovMLogPsi_vs_kg_myF}
\end{figure}

%
Coherent solutions for $0.028 \leq \kappa_g\leq 0.033$ are obtained using the 96-bit precision code,
and the radius, mass, charge, and central lapse can be observed  in Figure \ref{fig:RMQL_vs_U0_HighKappa}.  
The radius, mass, and charge appear to increase significantly as a function of $\kappa_g$, and the
central lapse again demonstrates significant gravitational time dilation relative to spatial infinity, 
$0.001 \lesssim \alpha_0 \lesssim 0.008$.
Figure \ref{fig:Schwarz_vs_U0_HighKappa} shows that for solutions with 
$0.029\lesssim\kappa_g\lesssim 0.033$, there are large regions of $u_0$-space that support superextremal 
solutions with radii on the order of $\kappa_g M_\infty$.  
These solutions are very dense objects that would form naked singularities if {runaway}
collapse were to occur.
Figure \ref{fig:RQovMLogPsi_vs_kg_myF} shows that with increasing gravitational coupling,
the radius of the coherent solutions increases dramatically and the required precision
to resolve the shooting solutions (indicated by $\log_{10}\left(\phi^+-\phi_0\right)$) also increases
rapidly.  
While the higher-precision code allows one to explore higher-energy (larger $\xi$,  $Q$, and $M_\infty$)
NTS solutions,
the apparent greater-than-exponential growth of the bubble radius as a function of $\kappa_g$ demands 
ever-increasing computational grid domains and the rapidly decreasing value of $\log_{10}\left(\phi^+-\phi_0\right)$
requires ever-greater numerical precision to fine-tune the initial shooting parameter ($\phi_0$).
As such, the work here stops with 96-bit precision but seems to definitively demonstrate the existence of 
superextremal ($|\Xi|>1$) coherent RN-AdS NTS solutions. 
%

\section{RN-A\lowercase{d}S NTS  Evolution and Stability  \label{sec:Evolution}}

While the solutions to
(\ref{eqn:CoherentBigPhi}),
(\ref{eqn:CoherentPhi}),
(\ref{eqn:CoherentEr}),
(\ref{eqn:CoherentU}), 
(\ref{eqn:CoherentA}), 
and
(\ref{eqn:CoherentAlpha})
satisfy the stationary coherent ansatz, they may not be stable to perturbations over time.
The long-term stability of the  coherent solutions is explored in this section by fully time-evolving 
(\ref{eqn:KG}),
(\ref{eqn:Maxwell1}),
(\ref{eqn:Maxwell2}),
(\ref{eqn:TimeDerivMetric}),
(\ref{eqn:TimeDerivK}), 
(\ref{eqn:Hamiltonian}), 
and 
(\ref{eqn:Momentum}) 
with a perturbed set of coherent initial data.
Again using polar-areal slicing, the time-dependent spherically symmetric metric is taken to be
\begin{equation}
ds^2 = -\alpha^2(t,r) dt^2 + a^2(t,r) dr^2 + r^2 \left( d\theta^2 + \sin^2\theta d\phi^2\right),
\end{equation}
which results in the following hyperbolic equations of motion: 
\begin{eqnarray}
\partial_t\Pi_1 &=& \frac{1}{r^2}\partial_r\left( \frac{\alpha}{a}r^2 \Phi_1 \right) 
- 2 q \left(A_t\Pi_2 -  \frac{\alpha}{a}A_r\Phi_2\right)  \\
&& 
+ \phi_1q^2\left( \frac{aA_t^2}{\alpha} - \frac{\alpha A_r^2}{a}\right)
-\alpha a \sum_n \alpha_n \phi_1 \phi_\rho^{2n-2}  \nonumber \\
\partial_t\Pi_2 &=& \frac{1}{r^2}\partial_r\left( \frac{\alpha}{a}r^2 \Phi_2 \right) 
+ 2 q \left(A_t\Pi_1 -   \frac{\alpha}{a}A_r\Phi_1\right)   \\
&&
+ \phi_2 q^2\left(\frac{aA_t^2}{\alpha} - \frac{\alpha A_r^2}{a}\right)
-\alpha a \sum_n \alpha_n \phi_2 \phi_\rho^{2n-2}  \nonumber\\
\partial_t\Phi_1 &=& \partial_r\left(\frac{\alpha}{a} \Pi_1\right) \\
\partial_t\Phi_2 &=& \partial_r\left(\frac{\alpha}{a} \Pi_2\right) \\
\partial_t\phi_1 &=& \frac{\alpha}{a} \Pi_1 \\
\partial_t\phi_2 &=& \frac{\alpha}{a} \Pi_2 \\
\partial_t\left( E_r \right) &=&
 q \frac{\alpha}{a} \left( \phi_2\Phi_1 - \phi_1\Phi_2\right) 
+ q^2 \phi_\rho^2 \frac{\alpha}{a} A_r  
\label{eqn:Max1_PA_Lorentz}\\
\partial_t\left( \frac{a}{\alpha} A_t \right) &=& \frac{1}{r^2}\partial_r\left(\frac{\alpha}{a} r^2  A_r\right)
\label{eqn:Max3_PA_Lorentz}\\
\partial_t A_r  &=& \partial_r A_t - \alpha a E_r  \\
\partial_ta &=& -4 \pi \kappa_g \alpha a r j_r
\end{eqnarray}
and the following elliptical equations:
\begin{eqnarray}
\frac{a'}{a} &=& \frac{1- a^2}{2r} + 4\pi \kappa_g r a^2 \rho  \label{eqn:HamConstraint}\\
\frac{\alpha'}{\alpha} &=& \frac{a^2 - 1}{2r} + 4\pi\kappa_g r a^2 {S^r}_r \label{eqn:LapseEquation}\\
\hspace{-5mm}
\frac{1}{r^2} \partial_r\left( r^2E_r \right) &=&
J^t,
\label{eqn:Max2_PA_Lorentz}\end{eqnarray}
where
\begin{eqnarray}
\rho &=&    
\frac{1}{2} E_r^2 + \frac{1}{2a^2}\left( 
\Phi_1^2 +\Phi_2^2 + \Pi_1^2 +\Pi_2^2\right)   \nonumber \\
&& 
+q \left[ \frac{A_t}{\alpha a}  \left( \phi_2\Pi_1 - \phi_1\Pi_2\right)  + 
             \frac{A_r}{a^2}  \left( \phi_2\Phi_{1} - \phi_1\Phi_{2}\right) \right] \nonumber\\
&&
 +\frac{1}{2} q^2 \phi_\rho^2 \left( \frac{A_t^2}{\alpha^2} + \frac{A_r^2}{a^2} \right) + V\\
{S^r}_r &=& 
-\frac{1}{2} E_r^2 + \frac{1}{2a^2}\left( 
\Phi_1^2 +\Phi_2^2 + \Pi_1^2 +\Pi_2^2\right) \nonumber \\
&&
+q \left[ \frac{A_t}{\alpha a}  \left( \phi_2\Pi_1 - \phi_1\Pi_2\right)  + 
             \frac{A_r}{a^2}  \left( \phi_2\Phi_{1} - \phi_1\Phi_{2}\right) \right] \nonumber\\
&&
 +\frac{1}{2} q^2 \phi_\rho^2 \left( \frac{A_t^2}{\alpha^2} + \frac{A_r^2}{a^2} \right) - V \\
j_r &=& 
-\frac{q}{\alpha}A_t \left( \phi_2\Phi_1 - \phi_1\Phi_2\right)
-\frac{q}{a}A_r \left( \phi_2\Pi_1 - \phi_1\Pi_2\right)  \nonumber \\
&&
-\frac{1}{a}\left( \Pi_1\Phi_1 + \Pi_2\Phi_2 \right) 
-\frac{q^2}{\alpha}\phi_\rho^2 A_t A_r \\
J^t &=&  q \left( \phi_2\Pi_1 - \phi_1\Pi_2\right) 
+  q^2 \phi_\rho^2  \frac{aA_t }{\alpha}  
\end{eqnarray}
and the charge and mass are conserved,
\begin{eqnarray}
Q(t) &=&   4 \pi \int_{0}^{r_b} dr r^2 J^t (t,r)  \\
M(t) &=& 4 \pi \int_{0}^{r_b}  dr r^2 \rho(t,r)  . 
\end{eqnarray}
%
%
%
%

As discussed in \cite{AlcubierreGonzalezSalgado2004}, 
if one based the stability solely on time-evolving the coherent initial data, the perturbation would be 
determined by the truncation errors of the coherent solutions. 
To have a more controlled parameterized perturbation, the scalar field initial data are taken to be
the following ``charge perturbed" values:  
\begin{eqnarray}
\phi(0,r)   &=&  \phi_c(r) \label{eqn:ChargePerturbation1} \\
\Pi(0,r)   &=&  i\left( \frac{a}{\alpha}\right)  \left( \omega_c + \Delta\omega \right) \phi_c(r), 
\label{eqn:ChargePerturbation2}
\end{eqnarray}
where $\phi_c(r)$ is a solution to the coherent equations obtained in Section \ref{sec:Coherent}, 
$\omega_c$ is the angular frequency of the coherent solution, 
and $\Delta\omega$ 
is an  arbitrary perturbation.   
When $\Delta\omega=0$, the coherent initial data are unchanged. 
When using a nonzero $\Delta\omega$, the perturbation changes the rotation rate of the scalar field
in the  ($\phi_1$,$\phi_2$) internal space in the direction of the U(1) isometry, thereby directly increasing or 
decreasing the  $\partial_t$ component of the conserved Noether current (the charge).  
Since the perturbation changes both the charge and mass distribution of the spacetime, 
the electromagnetic and gravitational constraint equations are solved with the 
new $\phi(0,r)$ and $\Pi(0,r)$.
%

\begin{figure}[t]
\begin{center}
\hbox{
\includegraphics[width=8cm]{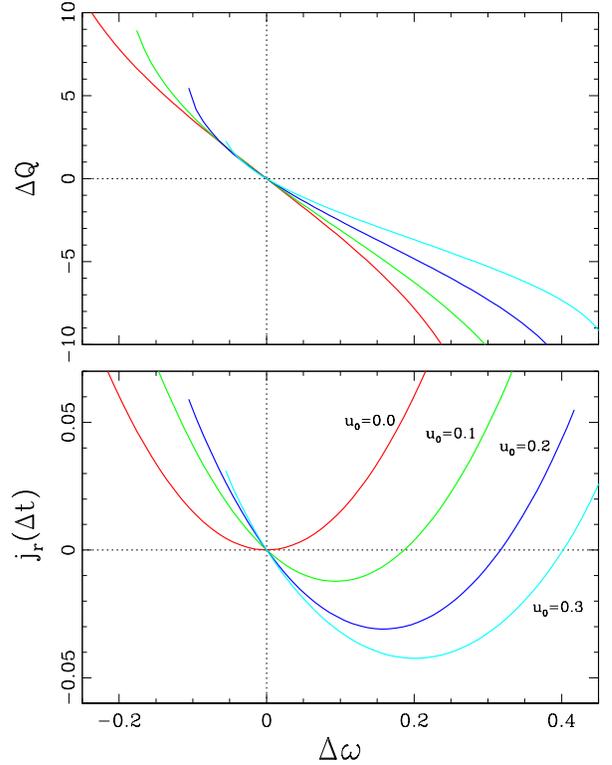}
}
\end{center}
\vspace{-10mm}
\caption{  
Plots of charge perturbation $\Delta Q$ and the radial 
ADM momentum $j_r(\Delta t)$ as a function of $\Delta \omega$, the
perturbation to the angular velocity of the complex scalar field. 
Plots shown are solutions for $\kappa_g = 0.011$ and $u_0 = \{ 0.0, 0.1, 0.2, 0.3\}$ in red, green,
blue, and cyan, respectively.  All solutions have $Q \geq 0$.
} 
\label{fig:dQ_jr_vs_dOmega}
\end{figure}

Figure \ref{fig:dQ_jr_vs_dOmega} demonstrates the effect that a given perturbation  has
on the charge and the radial component of the ADM momentum for a range of coherent solutions.  
Given the conventions 
used in this paper for the gauge covariant derivative, 
a negative perturbation to the angular frequency results in a positive perturbation to the charge of the
coherent bound state (top plot in Figure \ref{fig:dQ_jr_vs_dOmega}).  
The nature of the perturbation given by (\ref{eqn:ChargePerturbation1}) and (\ref{eqn:ChargePerturbation2}) 
is such that the geometric variables and 
their time derivatives are zero at $t=0$, but perturbations induce 
an imbalance of gravitational and electromagnetic forces.  
As such, $j_r$ is zero at $t=0$ but will be nonzero for $\Delta\omega\neq 0$ after one iteration forward in time 
and can serve as a measure of the effect the perturbation has on the dynamics of the NTS solution.
%
%
Looking at the bottom plot in Figure \ref{fig:dQ_jr_vs_dOmega}, one can see that since
the $u_0=0$ coherent solutions are charge-neutral, any perturbation creates a net
charge and thereby increases the Coulombic self-repulsion, resulting in outward radial motion ($j_r>0$)
independent of the sign of $\Delta\omega$.
On the other hand, since the coherent solutions with $u_0>0$ start with a positive charge before being perturbed, 
a negative $\Delta\omega$ perturbation implies a positive $\Delta Q$, 
and the amount of positive charge of the given solution is increased; the Coulombic self-repulsion therefore increases,
resulting in a radially outward motion, $j_r>0$. 
Conversely, $u_0>0$ coherent solutions with positive $\Delta\omega$ have negative $\Delta Q$, and 
the positive charge of the given solution is decreased;  the Coulombic self-repulsion therefore decreases,
resulting in a
radially inward motion, $j_r<0$.  With increasing $\Delta\omega$, enough negative charge can be added to 
make the solution charge-neutral (the minima of the curves); 
with additional $\Delta\omega$ (and more negative $\Delta Q$), 
the Coulombic self-repulsion starts increasing again, and 
the self-repulsion eventually balances out the gravitational attraction ($j_r=0$). 
With  even more negative $\Delta Q$, the solutions acquire enough negative charge to have 
net Coulombic repulsion ($j_r>0$). 
Equations (\ref{eqn:ChargePerturbation1}) and (\ref{eqn:ChargePerturbation2})  therefore serve as a simple 
parametric means to ``charge perturb" the coherent RN-AdS NTS solutions.

\subsection{Short-Term Stability }
\begin{figure}[t]
\newcommand{\picxyB}{8cm}
\includegraphics[width=\picxyB]{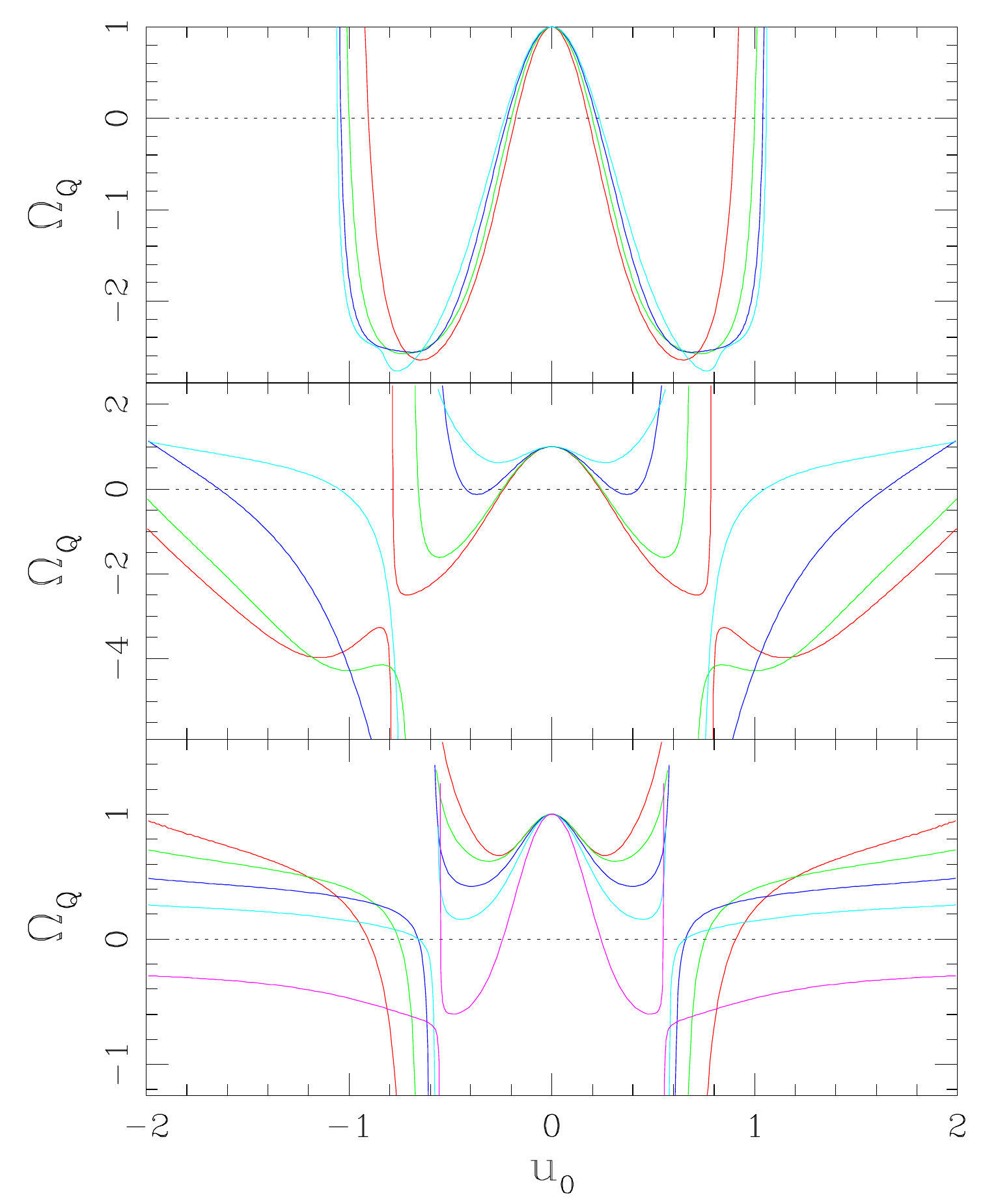}
\caption{  
Plots of $\Omega_Q$ as a function of $u_0$ for different values of $\kappa_g$.  
The top graph shows 
$\kappa_g = \{ 0.001, 0.008, 0.011, 0.01225\}$ in red, green,
blue, and cyan, respectively;
the middle graph shows 
$\kappa_g = \{ 0.0125, 0.013, 0.015, 0.020\}$ in red, green,
blue, and cyan, respectively;
the bottom graph shows 
$\kappa_g = \{ 0.022, 0.025, 0.028, 0.030, 0.032\}$ in red, green,
blue, cyan, and magenta, respectively.
Positive values of $\Omega_Q$ indicate unstable solutions, while negative values of $\Omega_Q$
indicate stable solutions.
}
\label{fig:dQdO_kGall}
\end{figure}

While the focus of this work is on the numerical time evolution of perturbed coherent solutions, 
it is useful to briefly consider the analytic stability of these solutions as well. 
It was shown in the context of scalar field Q-balls \cite{CorreiaSchmidt2001} 
(without a gauge field or coupling to gravity) that the quantity
\begin{equation}
\Omega_Q = \frac{\omega}{Q}\frac{\partial Q}{\partial\omega}
\label{eqn:StabilityCondition}
\end{equation}
determines the short-term stability of the soliton solution. For $\Omega_Q < 0$, Q-balls are stable, while
for $\Omega_Q >0$, they are unstable.  
This stability condition was observed to be true for RN-AdS solitons as well, but only in the gravitational
weak-field limit, which in this context can be considered to be for coherent 
solutions with $\alpha_0 \gtrsim 0.4$ and $\kappa_g \lesssim 0.015$.  
Figure \ref{fig:dQdO_kGall} shows $\Omega_Q$ as a function of $u_0$ for many values of gravitational
coupling.
Since $\Omega_Q(u_0=0)=1$ for all values of gravitational coupling, one can clearly see that 
charge-neutral solutions will be unstable.  
For $\kappa_g \lesssim 0.015$, there is a region in $u_0$-space  
where $\Omega_Q<0$ and the presence of a conserved charge leads to short-term stable solutions. 
%

\begin{figure}[t]
\newcommand{\picxyB}{8cm}
\includegraphics[width=\picxyB]{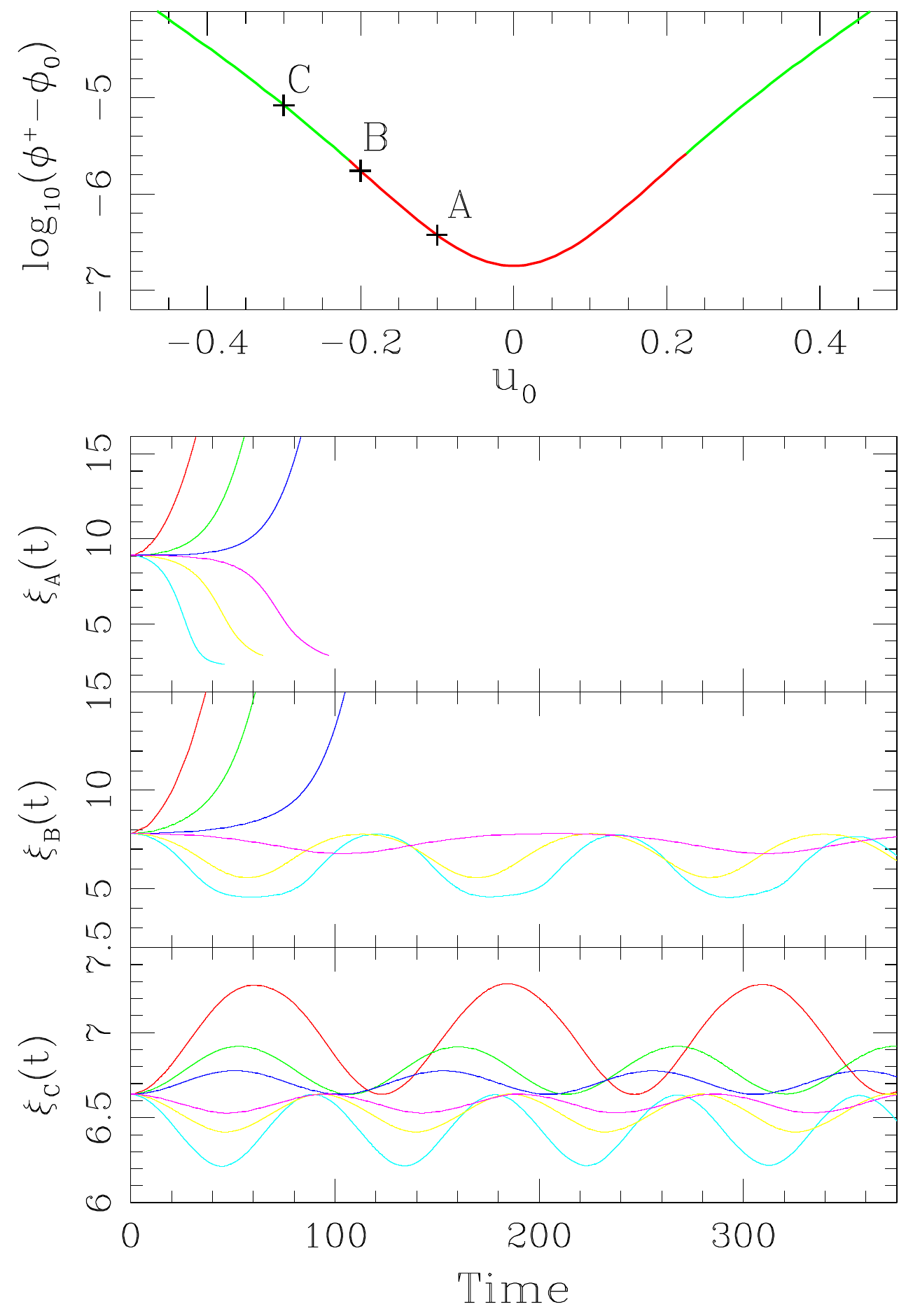}
\caption{  
Plots of $\xi(t)$ for $\kappa_g = 0.011$ for three different perturbed coherent solutions.  
The graphs of $\xi_A(t)$, $\xi_B(t)$, and $\xi_C(t)$ correspond to time evolutions of the 
perturbed coherent solutions at points A, B, and C, respectively, in the top graph.
The red, green, and blue curves represent time evolutions with a like-charged perturbation,
thereby increasing the Coulombic repulsion and resulting in an initial outward motion; 
the cyan, yellow, and magenta curves represent evolutions with opposite-charged perturbations
having the opposite effect.
The values of $\log_{10}(\phi^+ - \phi_0)$ are red for solutions with $\Omega_Q>0$ 
and green for solutions with $\Omega_Q<0$.
}
\label{fig:StabilityZones}
\end{figure}

\begin{figure}[t]
\newcommand{\picxyB}{8cm}
\includegraphics[width=\picxyB]{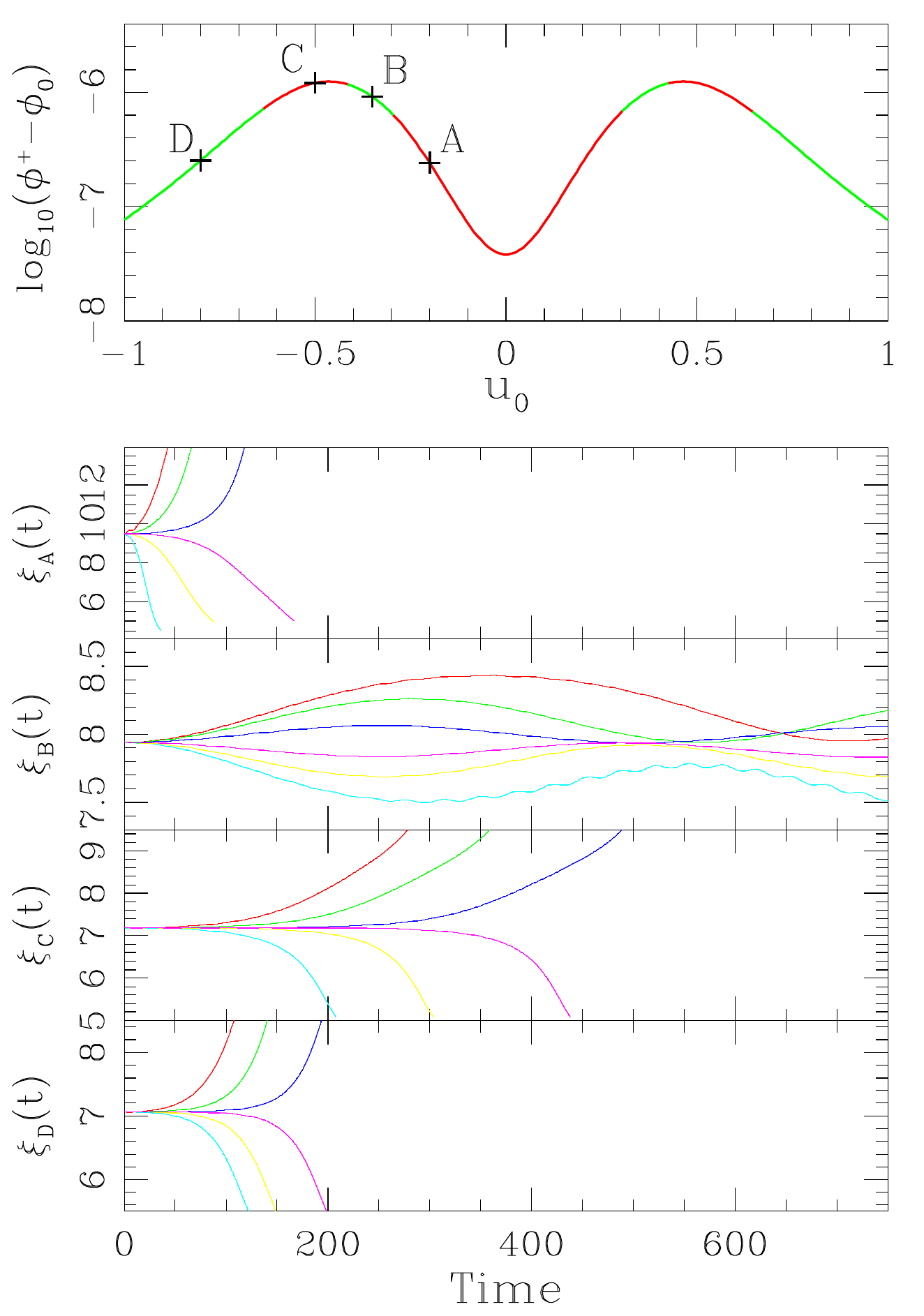}
\caption{  
Plots of $\xi(t)$ for $\kappa_g = 0.015$ for four different perturbed coherent solutions.  
The graphs of $\xi_A(t)$, $\xi_B(t)$, $\xi_C(t)$, and $\xi_D(t)$ correspond to time evolutions of the 
perturbed coherent solutions at points A, B, C, and D, respectively, in the top graph.
The red, green, and blue curves represent time evolutions with a like-charged perturbation,
thereby increasing the Coulombic repulsion and resulting in an initial outward motion; 
the cyan, yellow, and magenta curves represent evolutions with opposite-charged perturbations
having the opposite effect.
The values of $\log_{10}(\phi^+ - \phi_0)$ are red for solutions with $\Omega_Q>0$ 
and green for solutions with $\Omega_Q<0$.
Note that in case D, $\Omega_Q<0$ but the solutions are unstable.
}
\label{fig:StabilityZones2}
\end{figure}

%
Figure \ref{fig:StabilityZones} shows the time evolution of the radius, $\xi(t)$, 
for charge-perturbed initial data with $\kappa_g = 0.011$ for three different values of $u_0$ 
where the stability condition (\ref{eqn:StabilityCondition}) is observed to hold.
The top two $\xi(t)$ graphs show unstable time evolution in regions of $\Omega_Q>0$.  Perturbations 
that increase the net charge of the solution induce outward radial motion of the bubble wall due
to additional Coulombic self-repulsion; the location of the wall increases indefinitely and results 
in a bubble-induced phase transition to the true AdS vacuum.
Perturbations that decrease the magnitude of the charge of the solution result in an 
immediate collapse of the bubble wall that can have two different possible outcomes depending on the 
amount of charge.

Solutions with less charge ($Q<Q_{\rm TP}$) will collapse 
to within the RN outer horizon and form a black hole ($\xi_A(t)$).
Solutions with more charge ($Q>Q_{\rm TP}$) will collapse until the Coulombic 
self-repulsion causes the wall to ``bounce" back toward its original location ($\xi_B(t)$).
The point in $u_0$-space where $Q\approx Q_{\rm TP}$ is referred to as the ``triple point" because 
depending on the nature of the perturbation, the end state can be an RN-AdS black hole, 
a phase transition to the AdS true vacuum, or the false vacuum containing an RN-AdS bound state.
The bottom graph demonstrates the behavior of solutions in an $\Omega_Q<0$ region where the NTS
 solutions are stable to perturbations ($\xi_C(t)$).  
Similarly, Figure \ref{fig:StabilityZones2} shows the time evolution of the radius for different 
perturbed coherent solutions, this time with $\kappa_g = 0.015$.  
The stability condition holds for $|u_0| \lesssim 0.6$, but for $|u_0|\gtrsim 0.6$, solutions 
are gravitationally strong-field solutions ($\alpha_0 \lesssim 0.1$) and are unstable, 
even though there are regions where $\Omega_Q<0$.

\subsection{Long-Term Stability and Phase Diagrams}

Although $\Omega_Q$ can be an indicator of short-term stability under certain conditions, 
it cannot accurately predict the long-term fate of RN-AdS solitons.
The long-term behavior of these solutions can be systematically understood by time evolving 
the perturbed initial data and using the end states and exit conditions defined in 
Table \ref{table:ExitCriteria} to create ``phase diagrams."

Remembering that the coherent initial data are bubble solutions that separate an AdS true vacuum interior 
from a RN false vacuum exterior, 
a phase transition (PT) is determined to have occurred when most of the space within a given 
radius, $r_0$, is converted to the AdS true vacuum.  $r_0$ was chosen to be large compared 
to the initial bubble radius and such that all observed solutions with $\xi > r_0$ 
led to runaway expanding bubbles.
A solution is determined to have formed a black hole (BH) when a value of $2M(r)/r$ 
exceeds a threshold, $\delta$.
A solution is determined to have dispersed (D) when the scalar field is nowhere greater than or equal to half-way 
between the true and false vacuum for a period of time, $T_\textrm{disp}$. Such solutions were 
never seen to support bound states and always left the false vacuum intact.
Finally, when over a time, $t_\textrm{max}$, a solution does not induce a phase transition, form a black hole, 
or disperse, it is considered a bound state (BS). 
It should be noted that PT, BH, and D solutions are definitively observed and the exit criteria were chosen such that
one could be confident that the solution remained in that state for all future time.  
The bound-state solutions, on the other hand, are determined by default in that the solution is of finite extent for 
at least $t = t_\textrm{max}$; one cannot assume a BS solution will remain a BS solution for all time.  
%

\begin{table}[t]
\vspace{10mm}
\begin{tabular}{lcr}
\hline
\hline
End State & \hspace{25mm} &  Condition \\
\hline
Phase Transition (PT) & & $\chi(\phi_\rho,r_0) \geq  \chi_0$   \\
Black Hole (BH)  &  & $ \textrm{max}\left[ \left( \frac{2M}{r}\right)_i \right] \geq  \delta $\\
Disperal (D) & &  $T_{\xi=0} \geq T_\textrm{disp} $\\
Bound State (BS) & & $t \geq t_\textrm{max}$ \\
\hline
\end{tabular}
\caption{Table of conditions that result in different end-states to the time evolution equations.
For most simulations used in this work, $\chi_0 = 0.9$, $\delta = 0.7$, $T_\textrm{disp} = 40$, 
and $t_\textrm{max}$ is set to approximately twice the size of the computational domain.
}
\label{table:ExitCriteria}
\end{table}

\begin{figure}[t]
\begin{center}
\hbox{
\includegraphics[width=8cm,height=8cm]{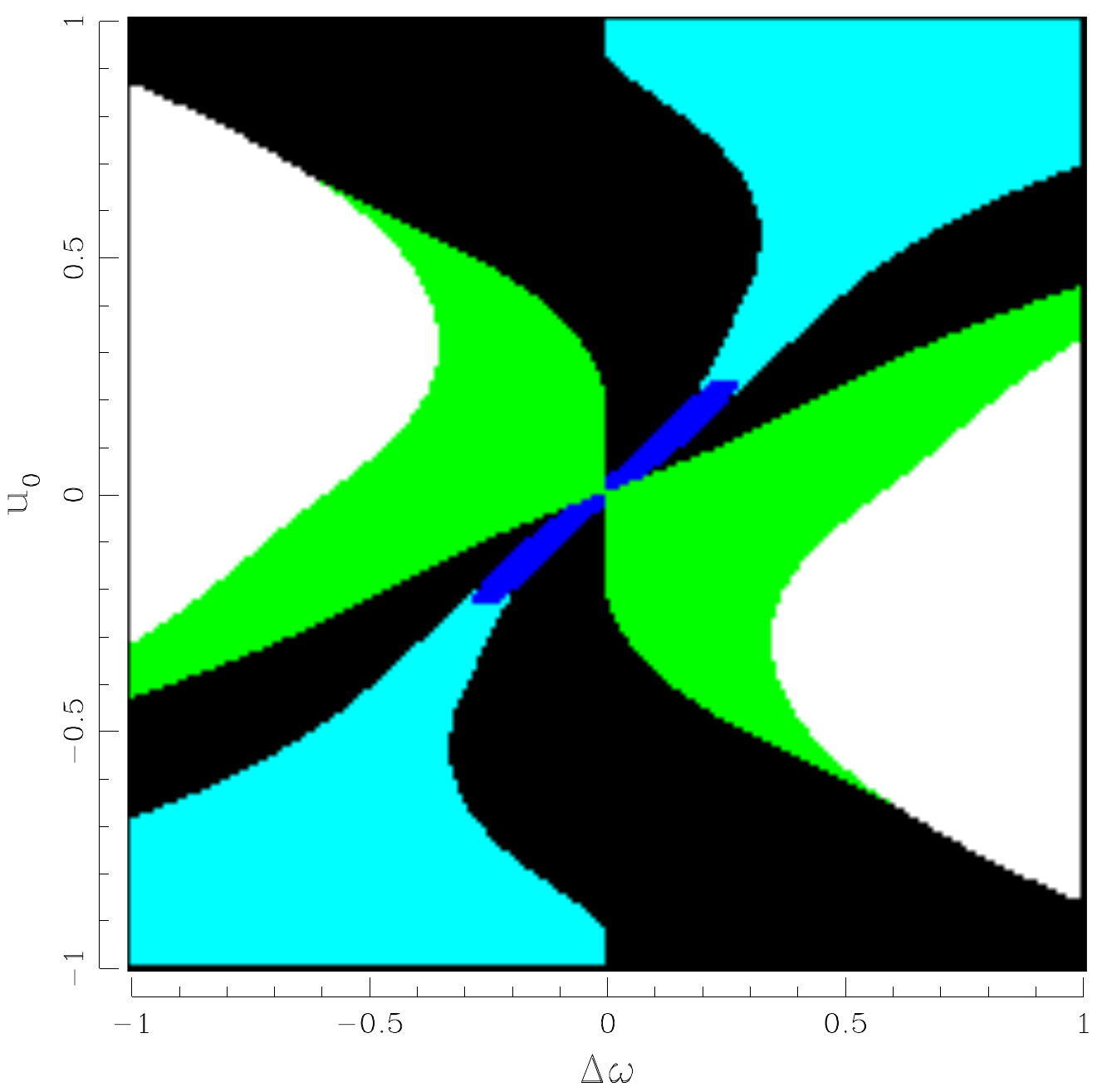}
}
\end{center}
\vspace{-10mm}
\caption{  
Bitmap of $(\Delta\omega,u_0)$ parameter space survey showing end states of 
time-evolved perturbed coherent solutions.  
The bitmap spans 
$\{\Delta\omega: -1 \leq \Delta\omega \leq 1\}$ on the horizontal axis and 
$\{u_0: -1 \lesssim u_0 \lesssim 1\}$ on the vertical axis
and is for gravitational coupling  $\kappa_g=0.003$.
The bitmap contains roughly 40,000 points, each the result of a time evolution with
an end state of BH (blue), PT (green), D (cyan), or BS
(black).  White points represent solutions that could not satisfy the $A_t(r\!\rightarrow\!\infty)=0$ boundary 
condition at $t=0$.
The existence of stable coherent bound states is demonstrated by black pixels along $\Delta\omega = 0$.
} 
\label{fig:BitmapExplanation}
\end{figure}

To illustrate the utility of the phase diagram approach, 
Figure \ref{fig:BitmapExplanation} shows the results from time-evolving charge-perturbed coherent solutions
with $\kappa_g = 0.003$.
Each pixel in the bitmap corresponds to a $(\Delta\omega,u_0)$ pair where a coherent solution with the 
given $u_0$ was perturbed by $\Delta\omega$ and time evolved until one of the conditions in Table 
\ref{table:ExitCriteria} was met.
For a coherent ($\Delta\omega=0$) solution to be considered stable, 
it must remain a bound state when subjected to both
positive and negative charge perturbations ($\Delta\omega <0$ and $\Delta\omega>0$, respectively).
For coherent solutions with $|u_0|\lesssim 0.2$, $\Omega_Q$ is positive and solutions 
are predicted to be unstable.
It can clearly be seen that solutions are indeed unstable to like-charged 
perturbations and lead to PT end states (green pixels).
Solutions with opposite-charged perturbations collapse but can lead to two different outcomes 
(BS or BH), depending on the amount of charge and the gravitational coupling.
For most values of $u_0$, the bubble walls begin to collapse, but the Coulombic self-repulsion 
leads to a bounce before the wall collapses to within the outer RN horizon and the solutions are modulated 
bound states (black pixels).
For very small $u_0$, on the other hand, 
the bubble walls do collapse to within the outer RN horizon and form black holes (blue pixels).
For coherent solutions where $0.2\lesssim |u_0|\lesssim 0.9$, $\Omega_Q$ is negative and evolutions 
are predicted to be stable to perturbations.
It can be seen that in this region that both positive and negative charge perturbations lead to bound states 
for the duration of the simulation, thereby indicating stable coherent solutions.
For coherent solutions with $|u_0| \gtrsim 0.9$, $\Omega_Q$ is positive and solutions are 
predicted to be unstable again; however this time, 
collapsing solutions end in dispersal and expanding solutions lead to modulated bound states 
without enough Coulombic self-repulsion to overcome the combined gravitational attraction and  
bubble wall surface tension.  

\begin{figure}[t]
\newcommand{\picxy}{2.5cm}
\vspace{5mm}
\begin{center}
\vbox{
\includegraphics[width=\picxy,height=\picxy]{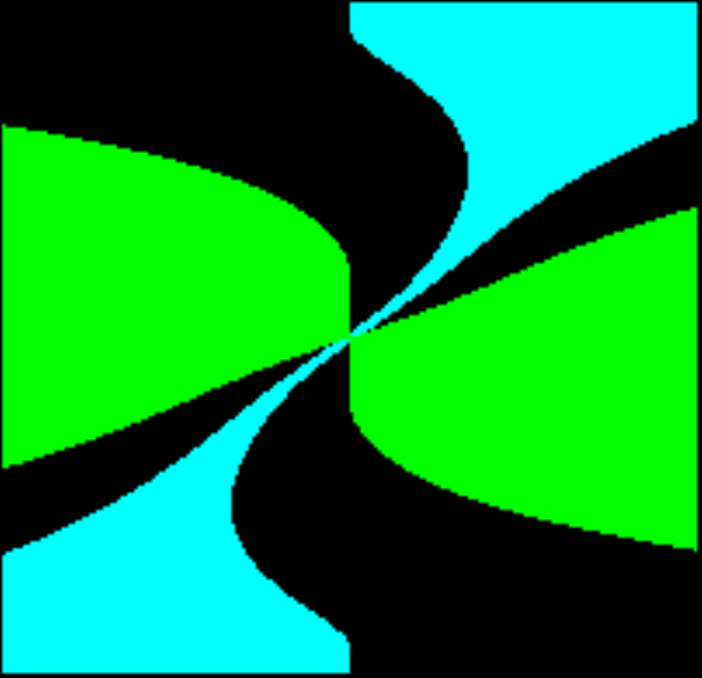}
\includegraphics[width=\picxy,height=\picxy]{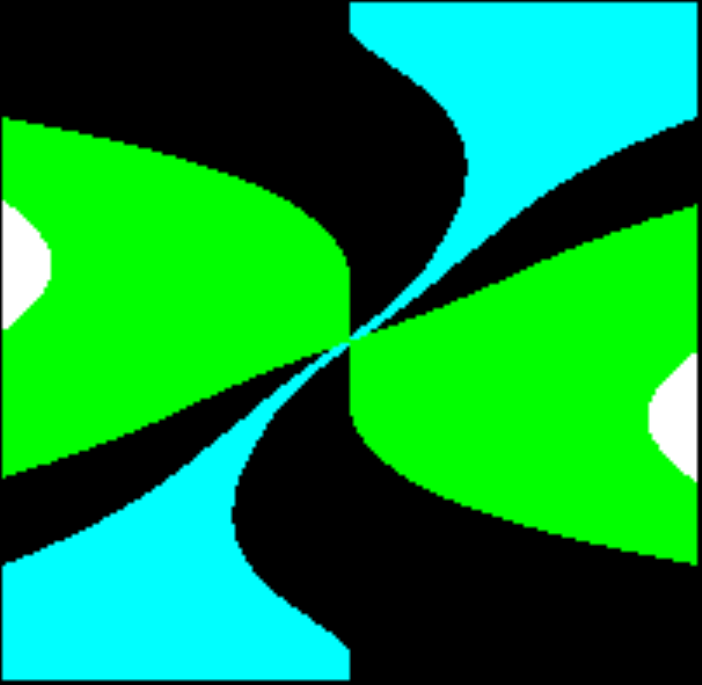}
\includegraphics[width=\picxy,height=\picxy]{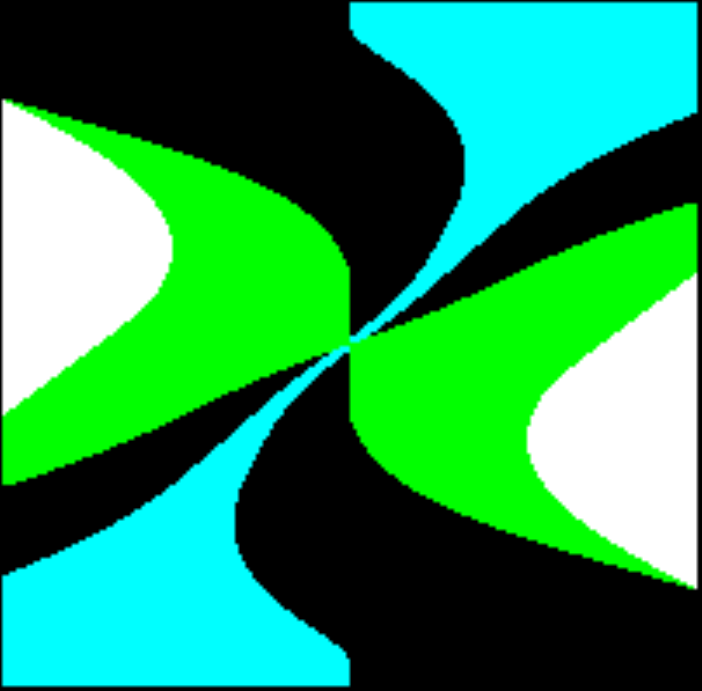}
}
\vspace{1mm}
\vbox{
\includegraphics[width=\picxy,height=\picxy]{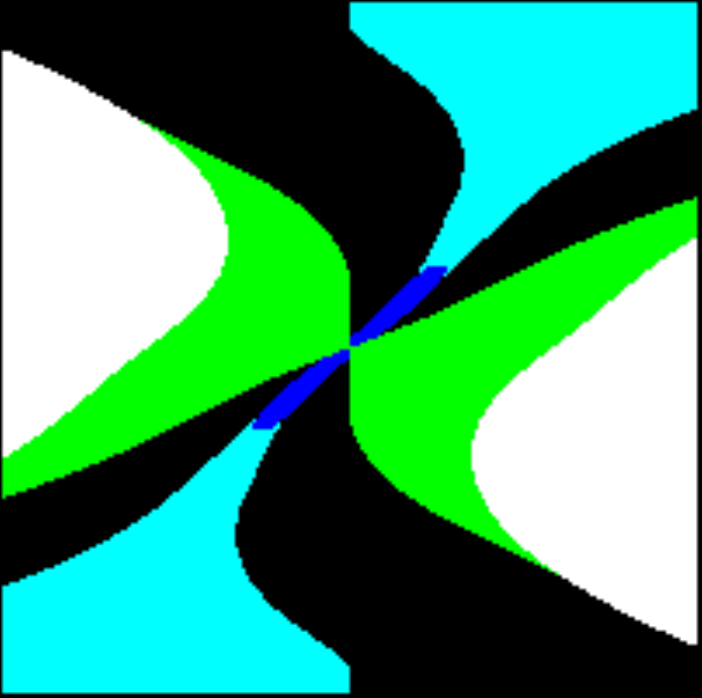}
\includegraphics[width=\picxy,height=\picxy]{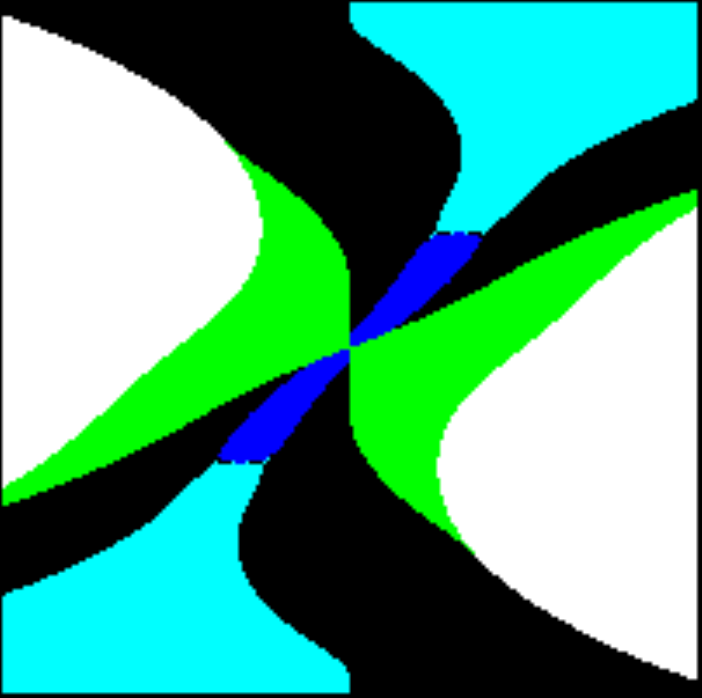}
\includegraphics[width=\picxy,height=\picxy]{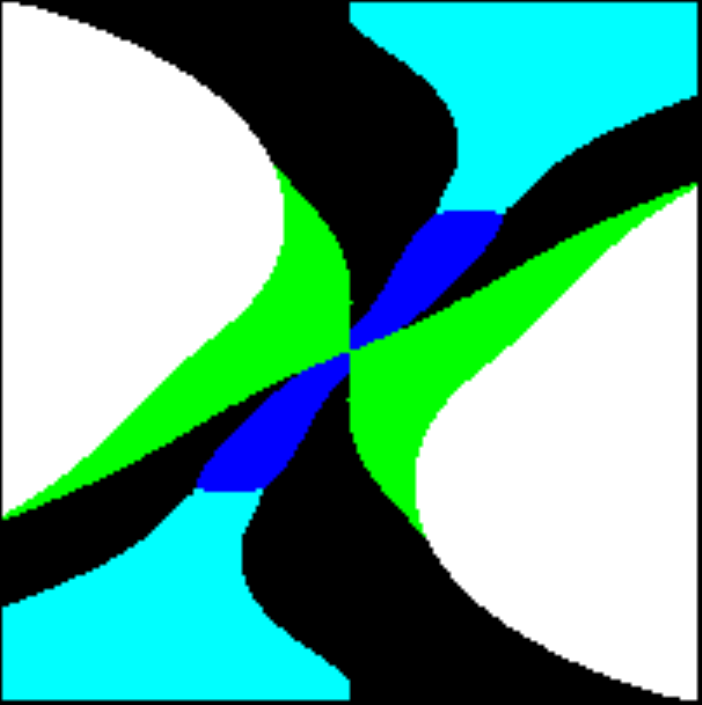}
}
\vspace{1mm}
\vbox{
\includegraphics[width=\picxy,height=\picxy]{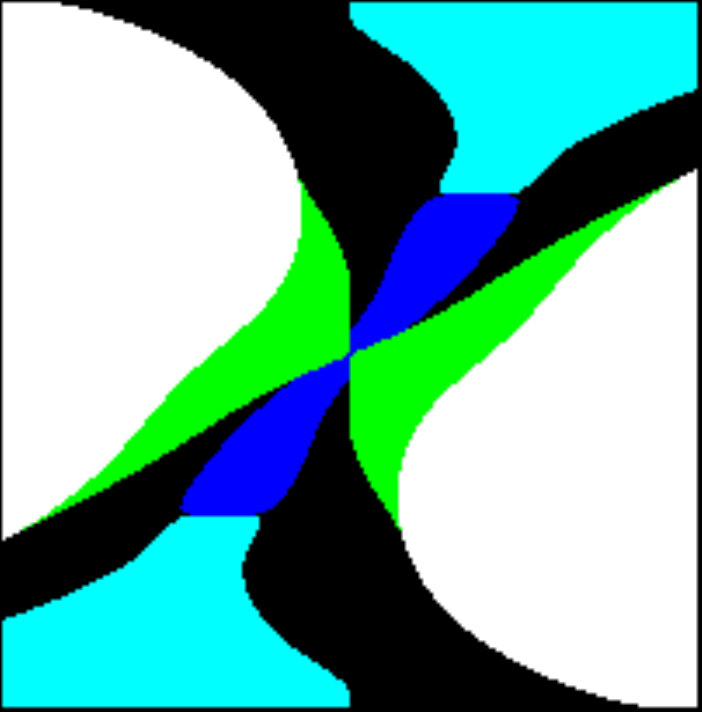}
\includegraphics[width=\picxy,height=\picxy]{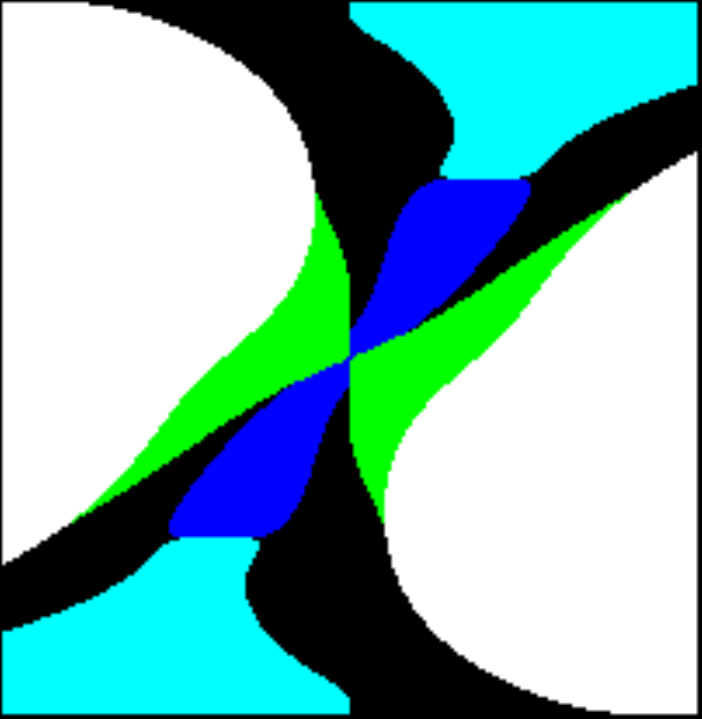}
\includegraphics[width=\picxy,height=\picxy]{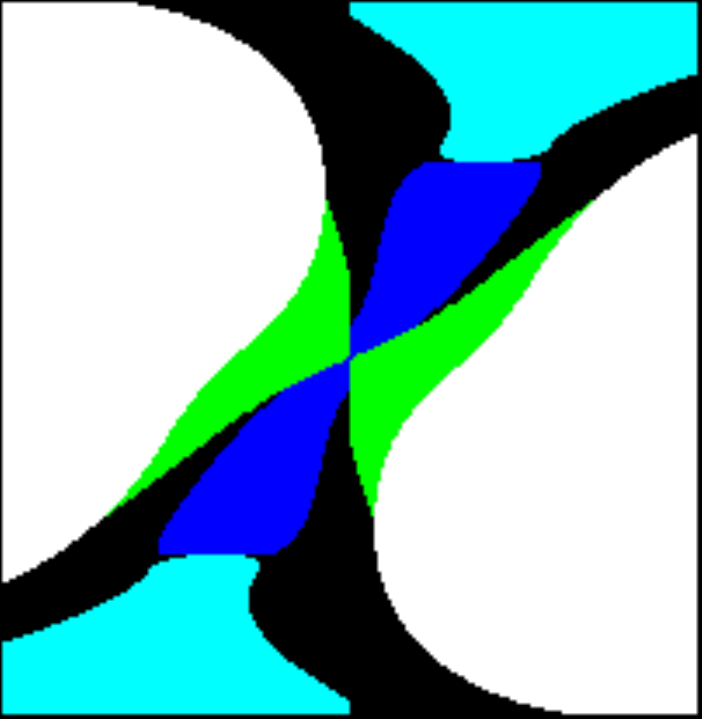}
}
\vspace{1mm}
\vbox{
\includegraphics[width=\picxy,height=\picxy]{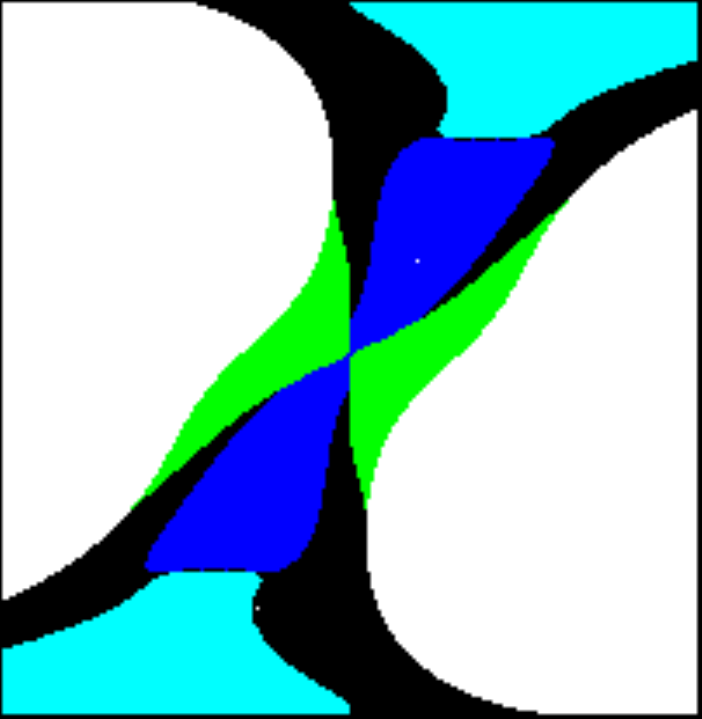}
\includegraphics[width=\picxy,height=\picxy]{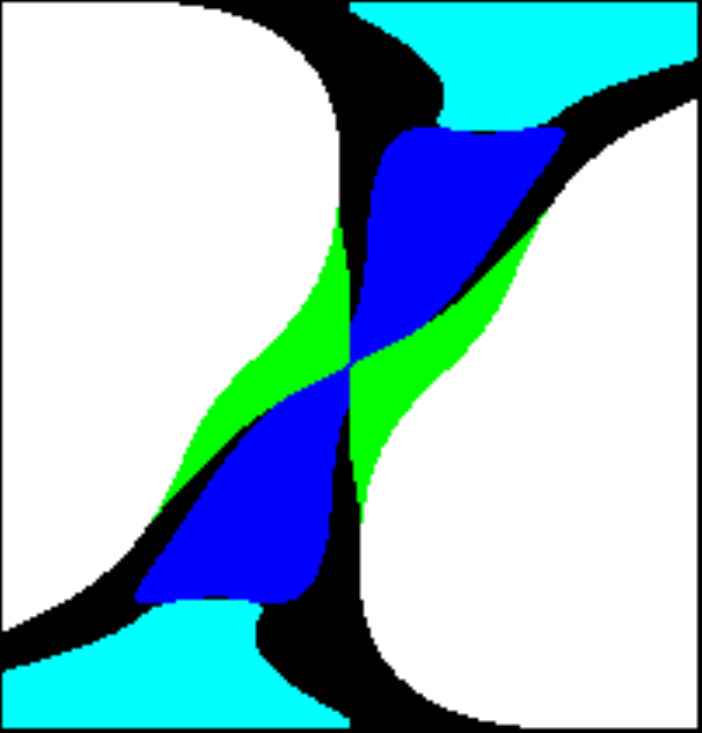}
\includegraphics[width=\picxy,height=\picxy]{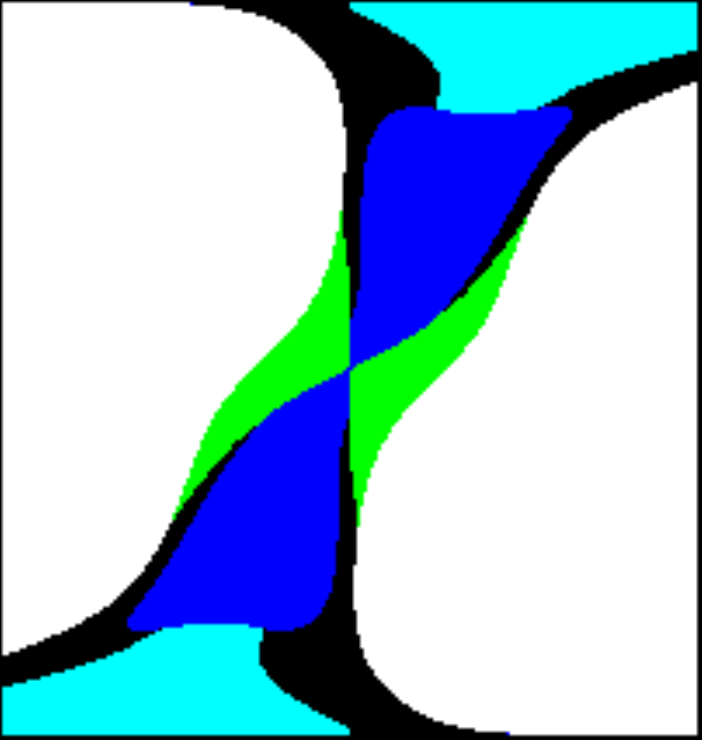}
}
\end{center}
\caption{  
Bitmaps of $(\Delta\omega,u_0)$ parameter space survey showing end states of 
time-evolved perturbed coherent solutions.  
The bitmaps span 
$\{\Delta\omega: -1 \leq \Delta\omega \leq 1\}$ on the horizontal axis and 
$\{u_0: -1 \lesssim u_0 \lesssim 1\}$ on the vertical axis. 
From left to right and top to bottom, the tiles represent different gravitational couplings
from $\kappa_g=0.0$ to $\kappa_g=0.011$, in steps of $\Delta\kappa_g=0.001$. 
Each bitmap contains roughly 40,000 points, each the result of a time evolution with
an end state of BH (blue), PT (green), D (cyan), or BS (black).  
White points represent solutions that could not satisfy the $A_t(r\!\rightarrow\!\infty)=0$ boundary 
condition at $t=0$.
The existence of stable coherent bound states is demonstrated by black pixels along $\Delta\omega = 0$.
}
\label{fig:KappaGBitmaps}
\end{figure}

\begin{figure}[t]
\newcommand{\picxyB}{4cm}
\vspace{1mm}
\hbox{
\includegraphics[width=\picxyB,height=\picxyB]{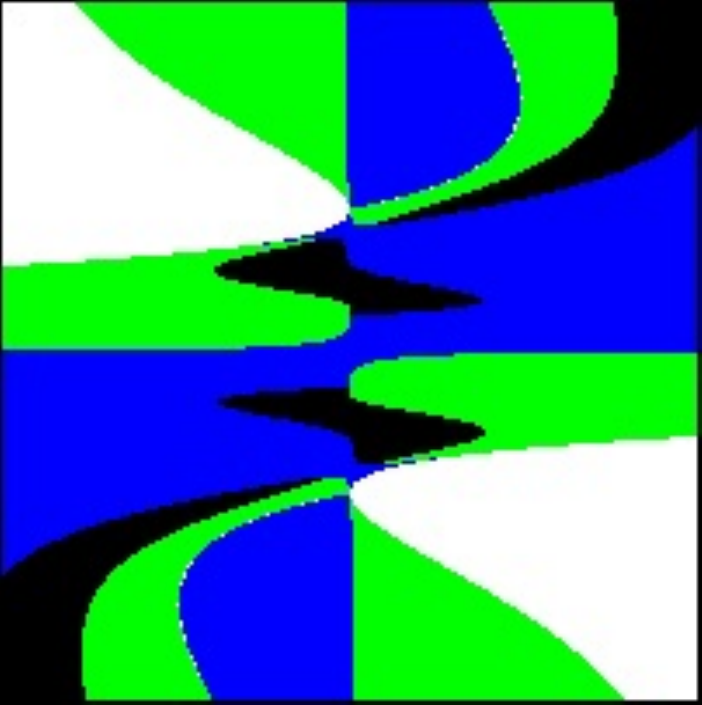}
\includegraphics[width=\picxyB,height=\picxyB]{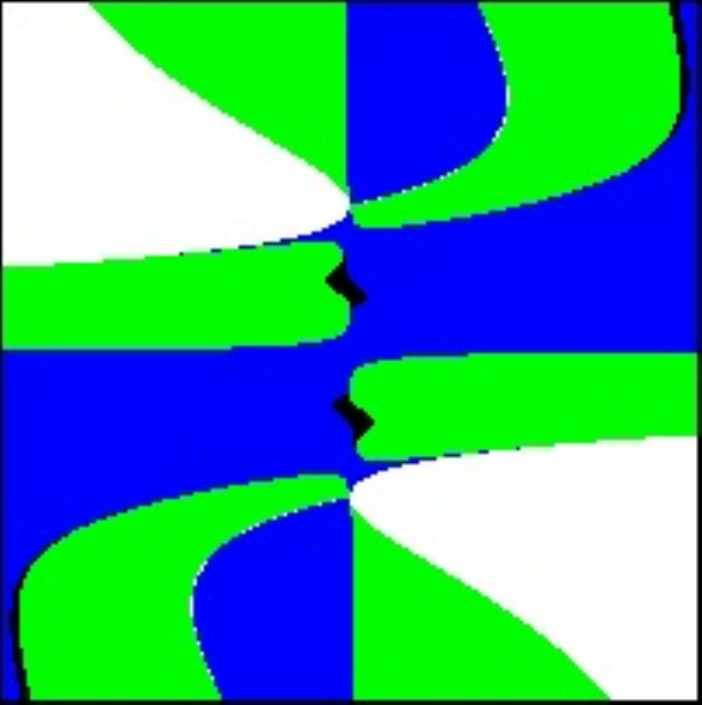}
}
\vspace{1mm}
\hbox{
\includegraphics[width=\picxyB]{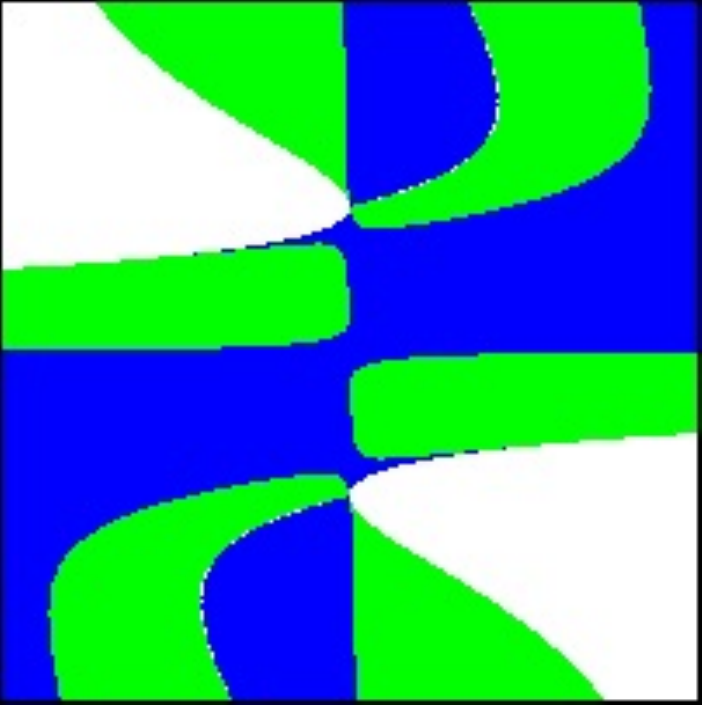}
\includegraphics[width=\picxyB]{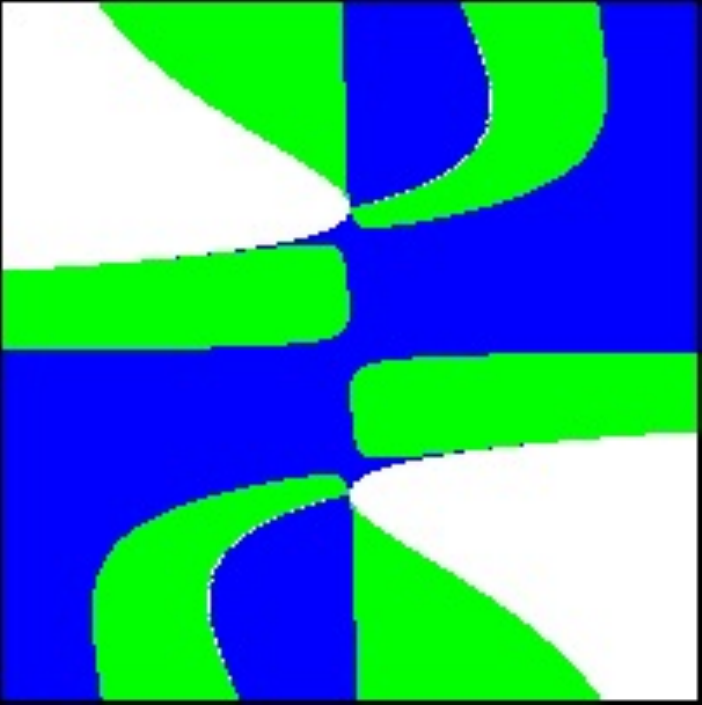} 
}
\caption{  
Bitmaps that span 
$\{\Delta\omega: -0.01 \leq \Delta\omega \leq 0.01\}$  on the horizontal axis and 
$\{u_0: -2 \lesssim u_0 \lesssim 2\}$ on the vertical axis
for $\kappa_g=\{0.014, 0.015, 0.016, 0.017\}$ 
(upper left, upper right, lower left, and lower right, respectively).
Each bitmap contains roughly 40,000 points, each the result of a time evolution with
an end-state of BH (blue), PT (green), D (cyan), or BS (black).  
White points represent solutions that could not satisfy the $A_t(r\!\rightarrow\!\infty)=0$ boundary 
condition at $t=0$.
The existence of stable coherent bound states is demonstrated by black pixels along $\Delta\omega = 0$.
}
\label{fig:KappaGBitmaps2}
\end{figure}

 Figure \ref{fig:KappaGBitmaps} shows twelve similar bitmaps, each covering the same range in 
 $(\Delta\omega,u_0)$ space but for different values of gravitational
 coupling.  With increasing gravitational coupling, more black holes appear, more 
 perturbed solutions do not have initial data that can satisfy the $A_t(r\rightarrow\infty)$ boundary
 condition, and the number of bound states seems to decrease.  
The fact that the area of solutions in $(\Delta\omega,u_0)$ space 
supporting stable bound states reduces dramatically with increasing
$\kappa_g$ is indicative of the lack of stable bound states for gravitationally strong-field  solutions.
Figure \ref{fig:KappaGBitmaps2} shows the transition across the $\kappa_g \approx 0.0155$ boundary,
above which no stable coherent bound states are observed. 
For these larger gravitational couplings, the range of $\Delta\omega$ values was adjusted to give insight 
into smaller perturbations, while a larger range of $u_0$-space was used to clearly demonstrate
the instability of solutions despite the $\Omega_Q < 0$ stability condition being met.

\begin{figure}[t]
\newcommand{\picxyB}{4cm}
\vspace{1mm}
\hbox{
\includegraphics[width=\picxyB,height=\picxyB]{kG_011.pdf}
\includegraphics[width=\picxyB,height=\picxyB]{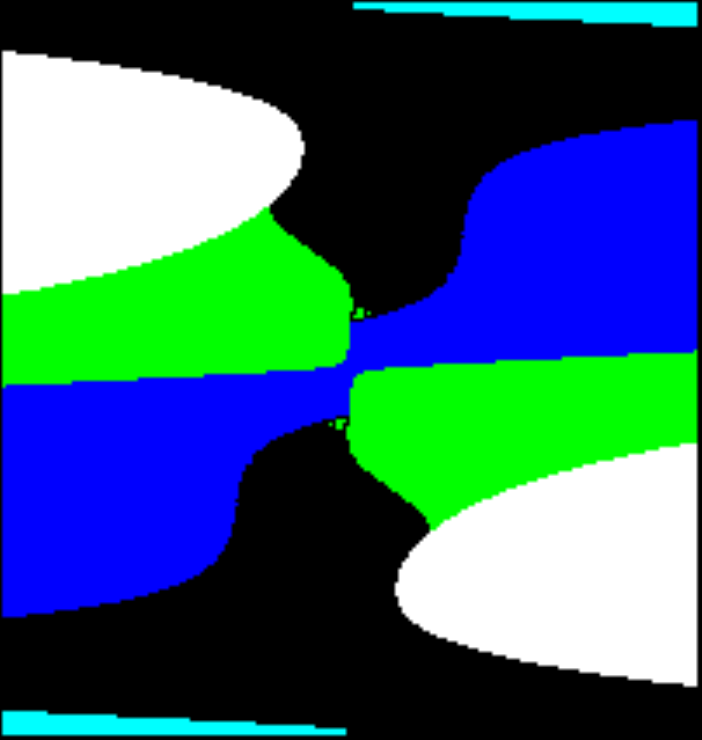}
}
\vspace{1mm}
\hbox{
\includegraphics[width=\picxyB]{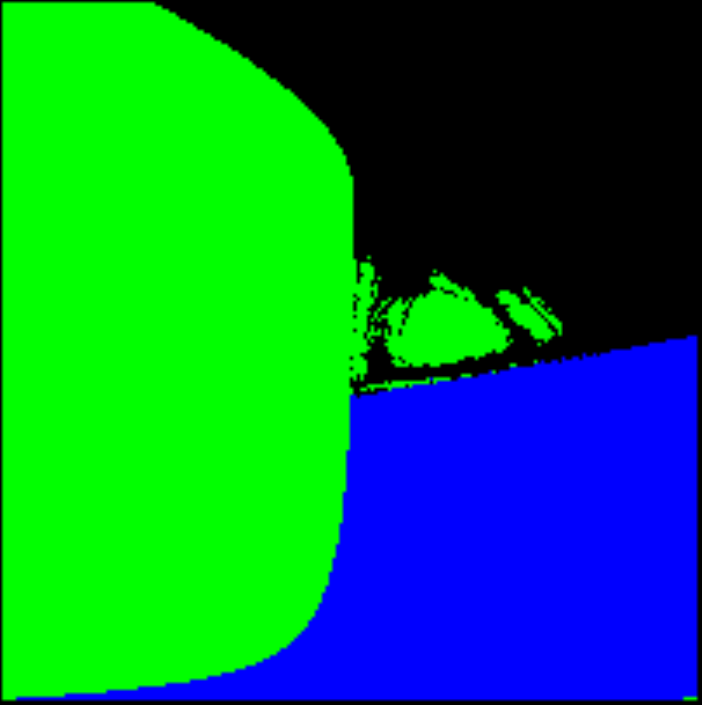}
\includegraphics[width=\picxyB]{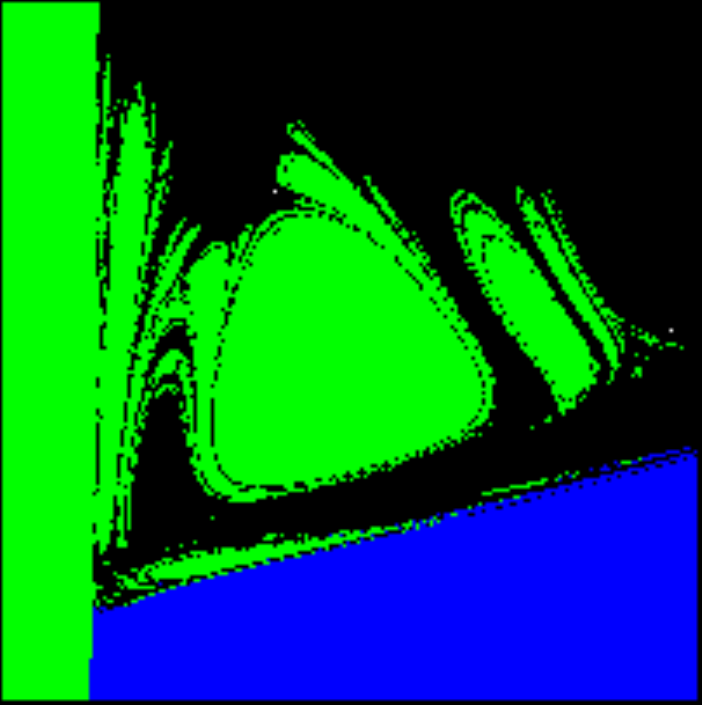} 
}
\caption{  
Bitmaps of successive magnification of $(\Delta\omega,u_0)$ parameter space survey showing end-states of 
time-evolved coherent solutions for $\kappa_g = 0.011$.
The upper-left bitmap spans 
$\{\Delta\omega: -1 \leq \Delta\omega \leq 1\}$  and 
$\{u_0: -1 \lesssim u_0 \lesssim 1\}$,
while the lower-right bitmap  spans
$\{\Delta\omega: -0.001 \leq \Delta\omega \leq 0.007\}$  and 
$\{u_0: 0.12 \lesssim u_0 \lesssim 0.20\}$.
Each bitmap contains roughly 40,000 points, each the result of a time evolution with
an end state of BH (blue), PT (green), D (cyan), or BS
(black).  White points represent solutions that could not satisfy the $A_t(r\!\rightarrow\!\infty)=0$ boundary 
condition at $t=0$.
}
\label{fig:BitmapsZoom}
\end{figure}

\begin{figure}[t]
\newcommand{\picxyB}{8cm}
\vspace{1mm}
\includegraphics[width=\picxyB]{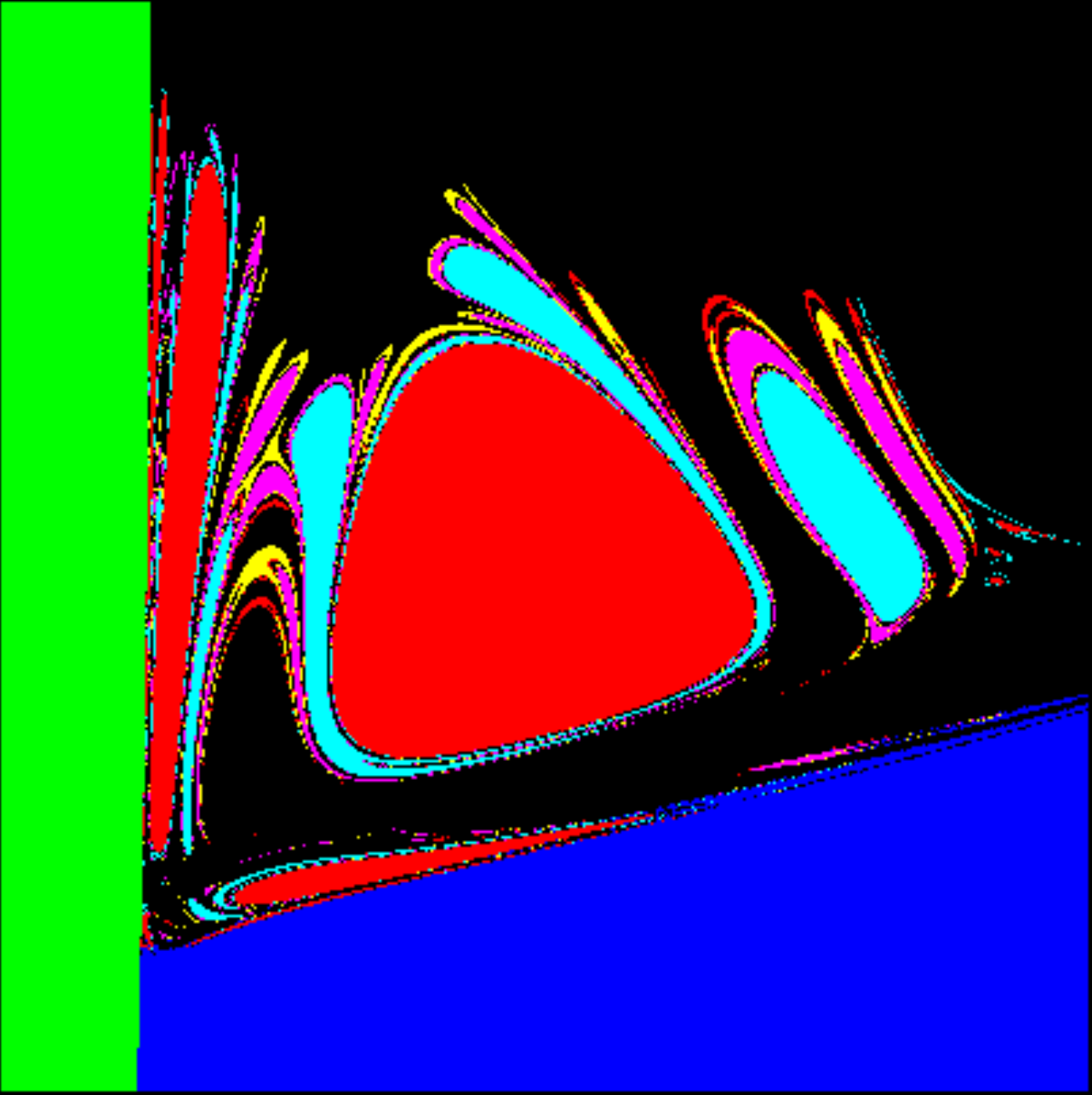}
\caption{  
Bitmap of $(\Delta\omega,u_0)$ parameter space survey showing end-states of 
time-evolved perturbed coherent solutions.  
The bitmap spans 
$\{\Delta\omega: -0.001 \leq \Delta\omega \leq 0.007\}$  on the horizontal axis and 
$\{u_0: 0.12 \lesssim u_0 \lesssim 0.20\}$ on the vertical axis
and is for gravitational coupling  $\kappa_g=0.011$.
The bitmap contains roughly 160,000 points, each the result of a time evolution with
an end state of BH (blue), immediate ($n_{\rm mod}=0$)
PT (green),  or BS (black).  
PT solutions with $n_{\rm mod}>0$ are colored based on their value of $n_{\rm mod}$ by cycling
through a color palette of red ($n_{\rm mod}=1$), cyan ($n_{\rm mod}=2$), magenta ($n_{\rm mod}=3$), 
and yellow ($n_{\rm mod}=4$); the colors repeat for $n_{\rm mod}>4$. 
}
\label{fig:Zoom3}
\end{figure}

\begin{figure}[t]
\newcommand{\picxyB}{8cm}
\vspace{1mm}
\hbox{
\includegraphics[width=\picxyB]{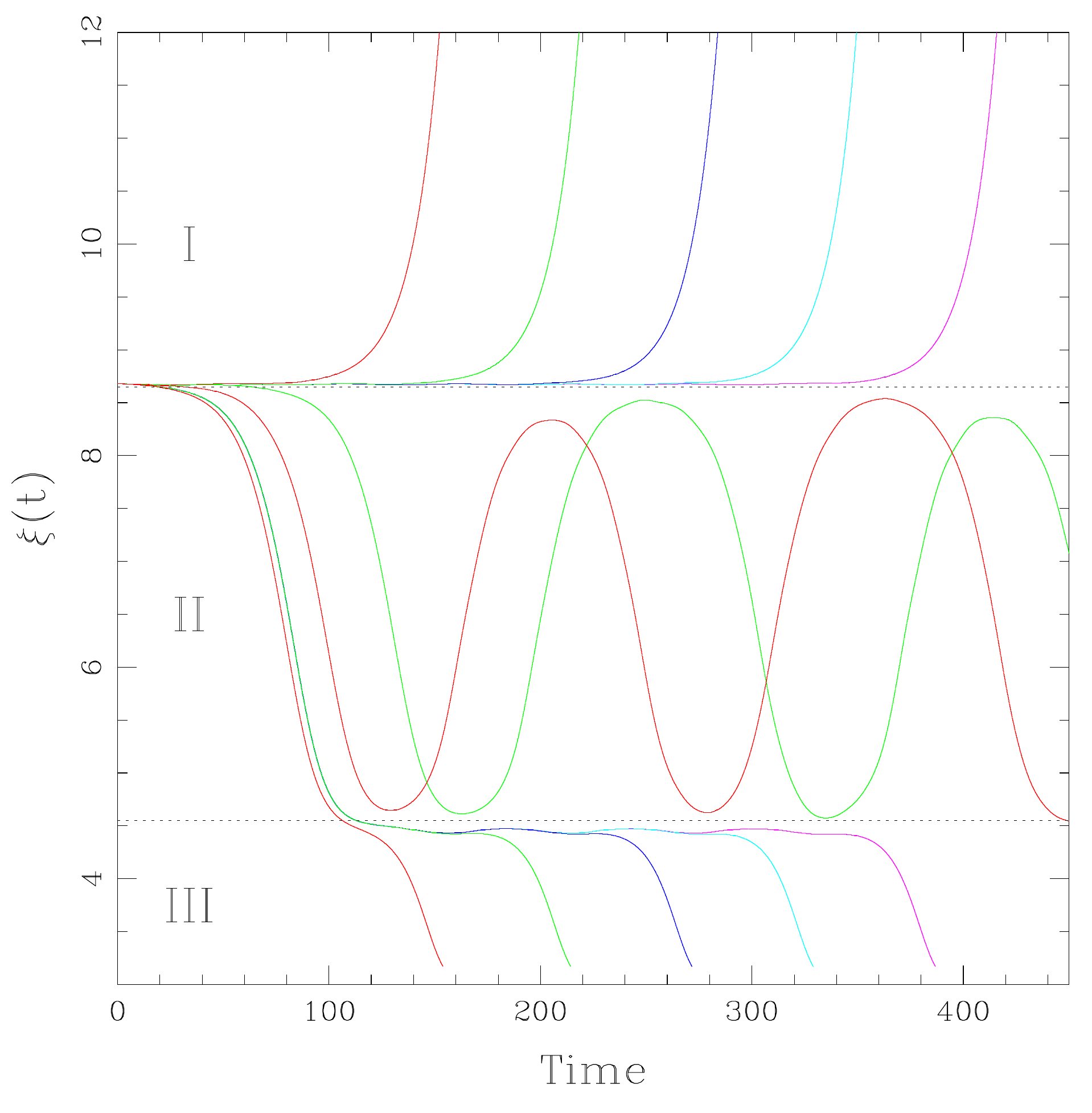}
}
\caption{  
Plots of $\xi(t)$ for solutions that induce phase transitions (region I above and solid green region in 
Figure \ref{fig:Zoom3}), 
solutions that form black holes (region III above and solid blue region in 
Figure \ref{fig:Zoom3}), 
and solutions that appear to create bound states (region II above and section between solid blue and green
regions in Figure \ref{fig:Zoom3}).
All solutions were generated by charge-perturbing the same $u_0= 0.1312$ coherent solution.}
\label{fig:PT_BH_Thresh}
\end{figure}

\begin{figure}[t]
\newcommand{\picxyB}{8cm}
\vspace{1mm}
\hbox{
\includegraphics[width=\picxyB]{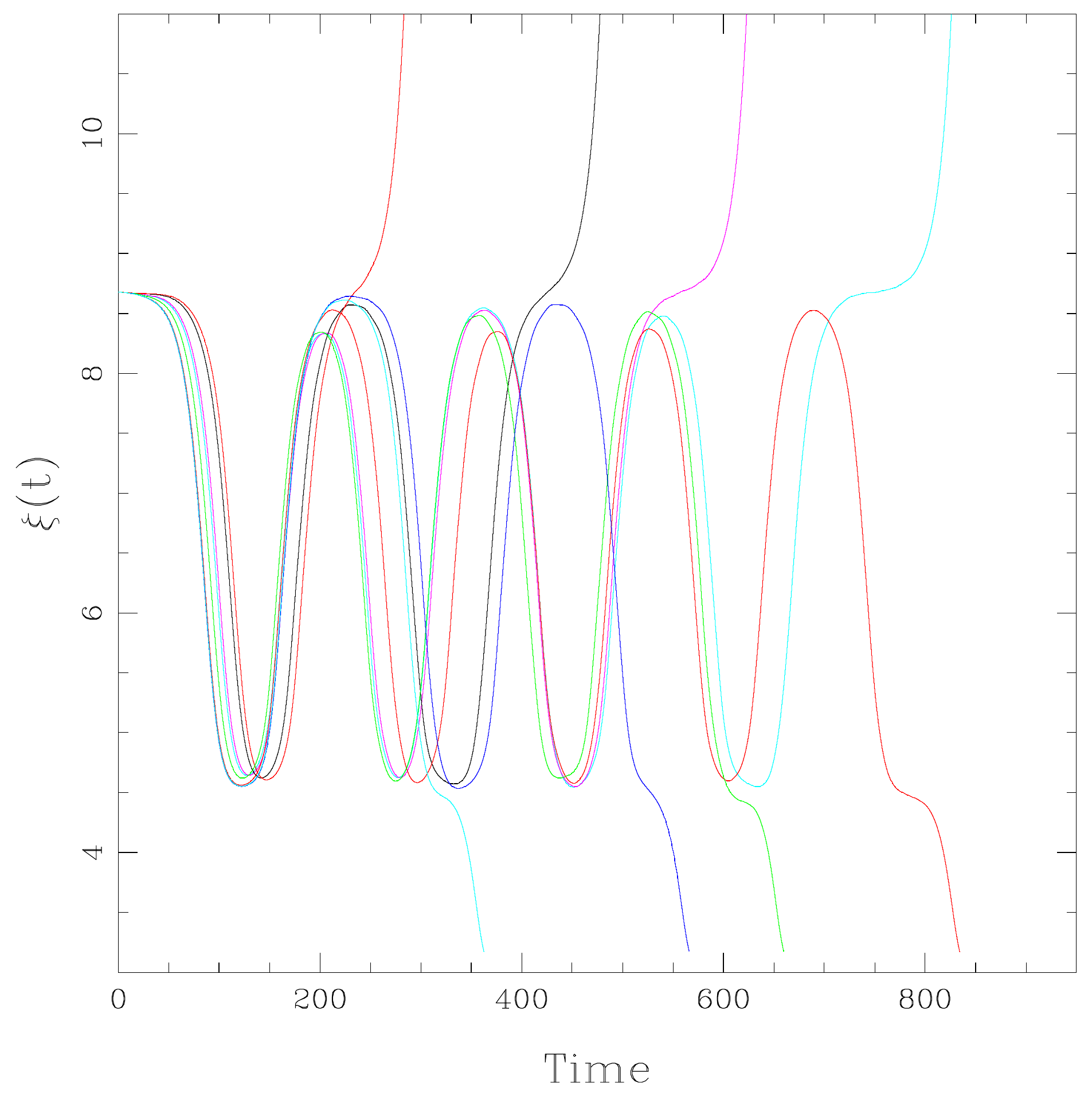}
}
\caption{  
Plots of $\xi(t)$ for solutions that induce phase transitions or create black holes 
after ``bouncing" for $\rm n_{mod} = 1,2,3,4$. 
All solutions were generated by charge-perturbing the same $u_0= 0.168$ coherent solution. 
The expanding PT solutions shown above correspond to $n_{\rm mod}>0$ colored pixels in 
Figure \ref{fig:Zoom3}.
}
\label{fig:PT_BH_nmods}
\end{figure}

Figure \ref{fig:BitmapsZoom} shows finer detail of the phase diagram for $\kappa_g = 0.011$.
The aforementioned ``triple point" can be seen around $u_0\approx 0.13$, where 
BH, PT, and BS solutions all exist in close proximity in 
$(\Delta\omega,u_0)$ space.
Of particular interest is that near the triple point, a fractal structure is observed in the phase diagram.
These solutions arise from ``bounce" solutions that initially collapse but eventually bounce back and 
expand enough to become runaway PT solutions.
Figure \ref{fig:Zoom3} shows these solutions colored based on the number of bounces (or modulations),
$n_{\rm mod}$,
they undergo before inducing a runaway phase transition.  This behavior is similar to the fractal boundary 
basins observed in real scalar field oscillon dynamics \cite{Honda2010}. 
Figures \ref{fig:PT_BH_Thresh} and \ref{fig:PT_BH_nmods} show the time evolution of the bubble radius 
for two different coherent solutions, $u_0=0.1312$ and $u_0=0.168$, for a variety of perturbations, $\Delta\omega$.  
Figure \ref{fig:PT_BH_Thresh} shows evolutions for three regions of the phase diagram. 
Region I is for $\Delta\omega<0$, where all solutions induce a phase transition (the green region 
on the left side of the bitmap in figure \ref{fig:Zoom3}).  
Region III is for larger $\Delta\omega$, where all solutions collapse to form black holes (the blue region 
on the lower-right side of the bitmap in figure \ref{fig:Zoom3}).
Region II is the intermediate region, where BS or bounce (PT or BH) solutions are supported.
Figure \ref{fig:PT_BH_nmods} shows time evolutions for bounce solutions like those in Region II of 
Figure  \ref{fig:PT_BH_Thresh}, but for $u_0=0.168$. 
Values of $\Delta\omega$ were chosen that demonstrate bounces prior to either inducing a phase transition
or collapsing to a black hole.  
%

\begin{figure}[t]
\newcommand{\picxyB}{8cm}
\vspace{1mm}
\hbox{
\includegraphics[width=\picxyB]{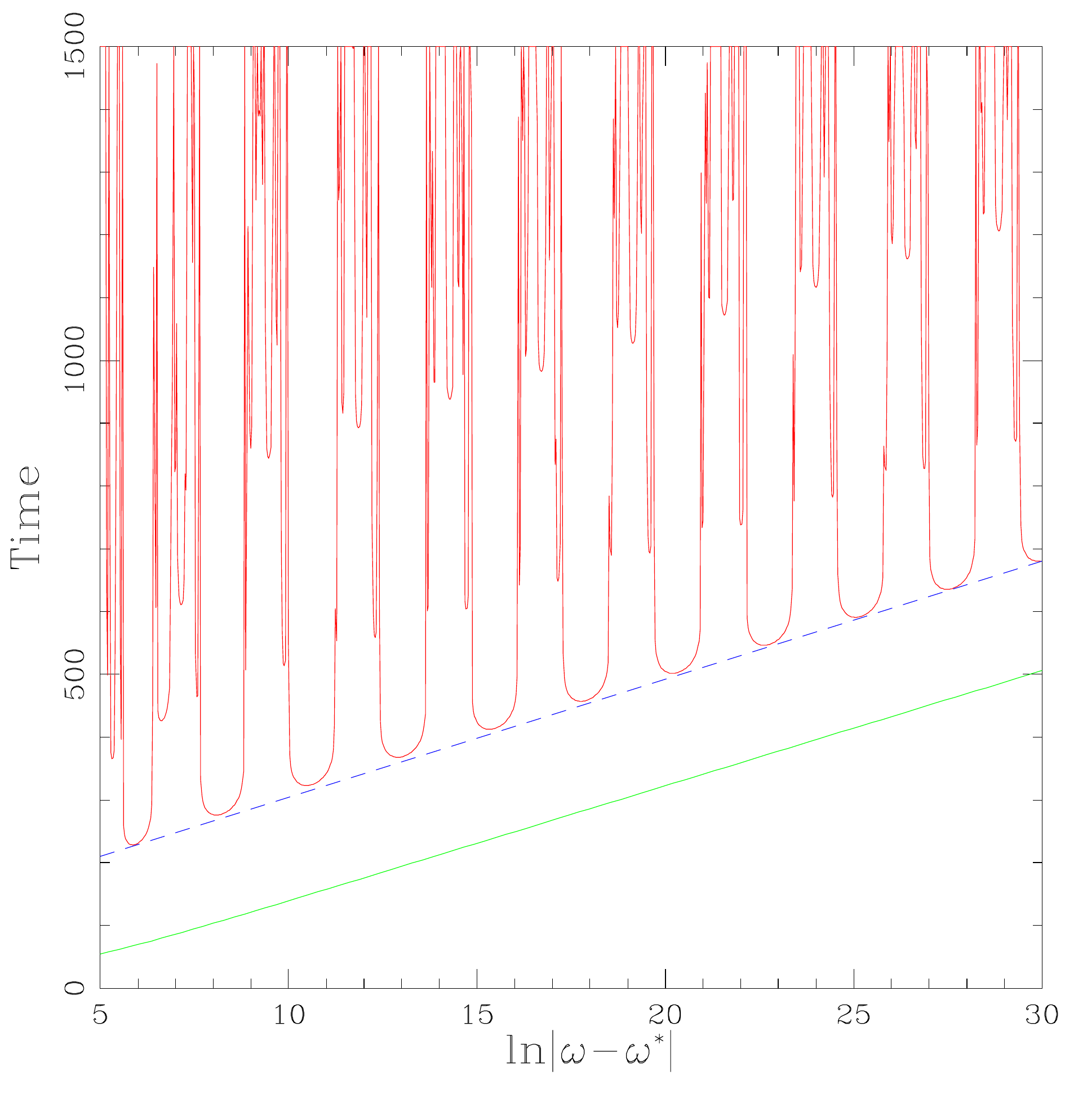}
}
\caption{  
Plot demonstrating the solution lifetime (before inducing a phase transition) as a function of the charge 
perturbation, $\omega$, as $\omega$ approaches the $n_{\rm mod}=0$ boundary for a $u_0= 0.168$ 
coherent solution.
The log-periodic nature of the solutions is apparent and the time-scaling exponent is measured to 
be $\gamma = 18.2$ on both sides of the threshold.  The green line is continuous since the perturbations
are clearly in the $n_{\rm mod}=0$ region, while the log-periodic bands have $n_{\rm mod}=1$ for the main lobes
outside the $n_{\rm mod}=0$ region.
}
\label{fig:CritSolnLarge}
\end{figure}

\begin{figure}[t]
\newcommand{\picxyB}{8cm}
\vspace{1mm}
\hbox{
\includegraphics[width=\picxyB]{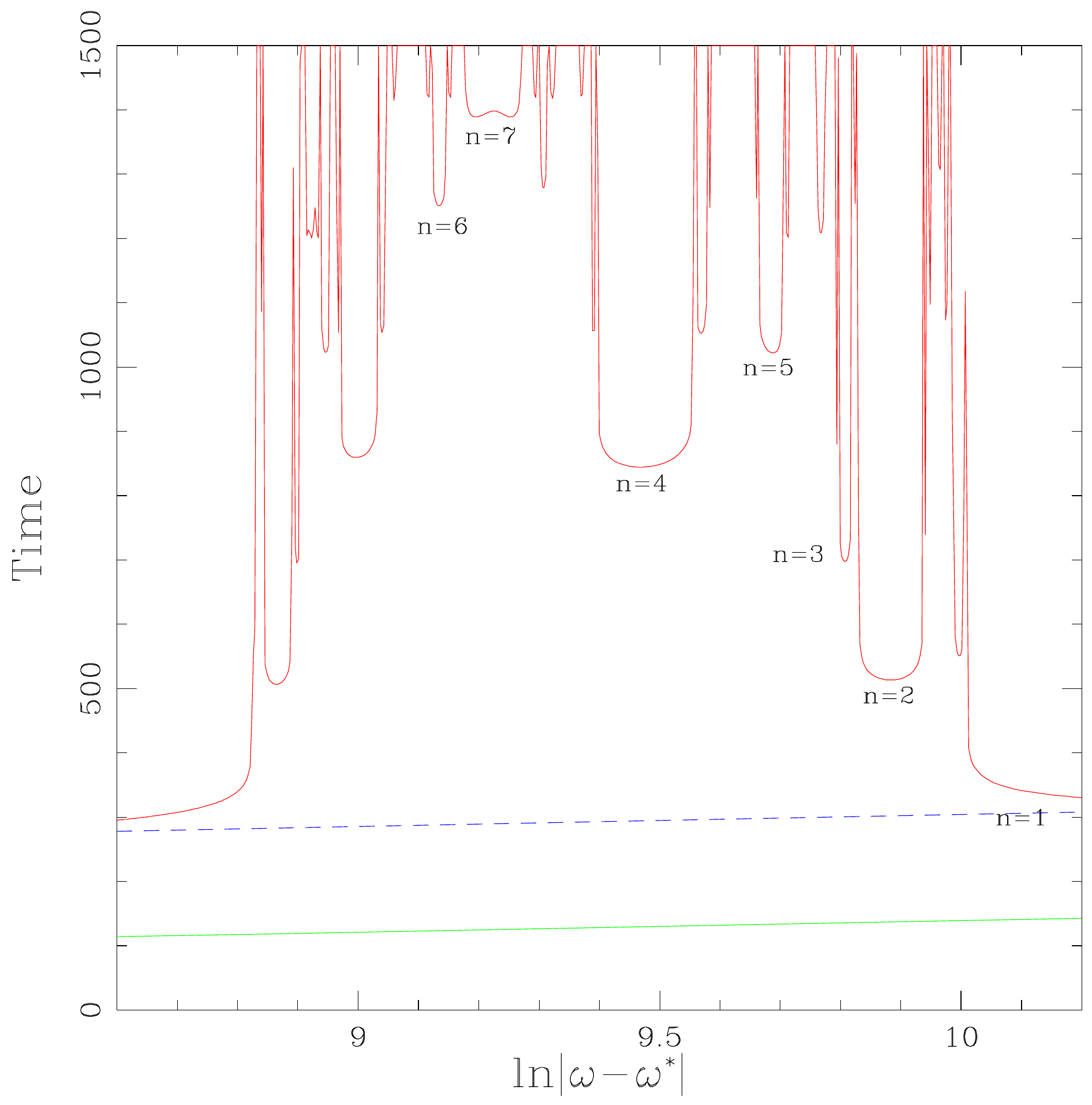}
}
\caption{  Plot of a zoomed-in region of Figure \ref{fig:CritSolnLarge} that shows the region between the 
$n_{\rm mod}=1$ lobes that still induce a phase transition, but with $n_{\rm mod}>1$.
}
\label{fig:CritSolnZoom}
\end{figure}
\begin{figure}[t]
\newcommand{\picxyB}{8cm}
\vspace{1mm}
\hbox{
\includegraphics[width=\picxyB]{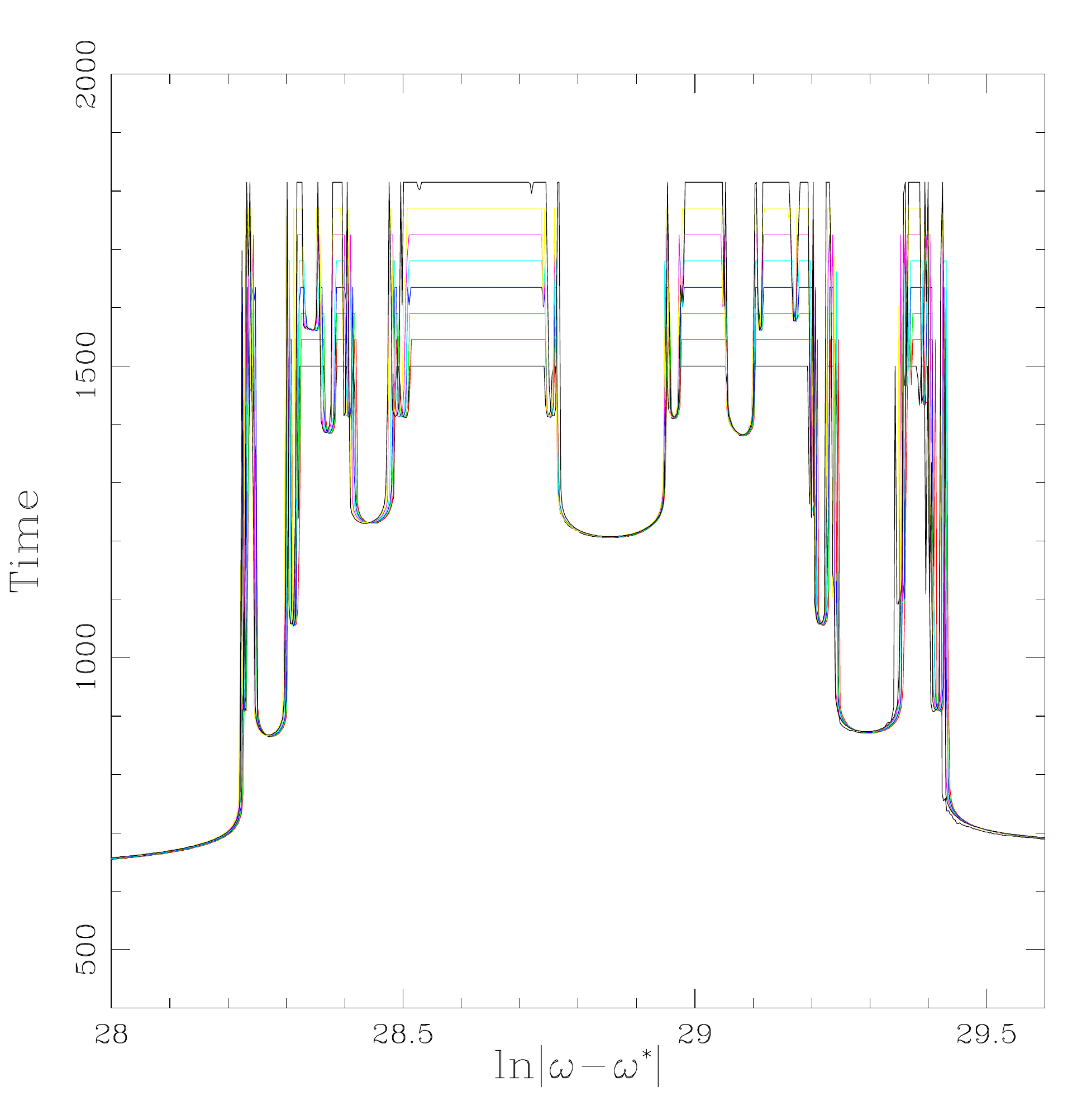}
}
\caption{  
Plot demonstrating the log-periodic nature of the PT solutions by overlaying 
$T(\ln|\omega-\omega^*|)$  with 
itself after offsetting by $\Delta_\omega = 2.4285$ and $\Delta_T = 45$.
$T(\ln|\omega-\omega^*| + n\Delta_\omega) + n\Delta_T$ has been plotted for 
$n=0,1,2,3,4,5,6,\rm{\ and \ } 7$ 
in black, yellow, magenta, cyan, blue, green, red, black, respectively.
In addition to the main $n_{\rm mod}=1$ lobes, the finer detail  
of the $n_{\rm mod}>1$ solutions also appears to be self-similar.
}
\label{fig:CritSolnOverlay}
\end{figure}

Similar to behavior observed in oscillon dynamics \cite{Honda2010}, PT regions of $(\Delta\omega,u_0)$
space with $n_{\rm mod}$ modulations are surrounded by PT regions with  
$(n_{\rm mod}+1)$ modulations that approach the region with $n_{\rm mod}$ modulations 
in a log-periodic fashion.  
To demonstrate the log-periodic nature of the bounce regions, one first needs to find the
boundary of such a region.  Since a PT region with $n_{\rm mod}$ modulations is surrounded
by either BS solutions or PT solutions with more than $n_{\rm mod}$ modulations, 
it is straightforward to vary $\omega$ and bisect on the boundary of the $n_{\rm mod}$ region, 
which is denoted $\omega^*$.
Figure \ref{fig:CritSolnLarge} shows the result of bisecting on a boundary of a $n_{\rm mod} = 0$ region.
Approaching $\omega^*$ from within the $n_{\rm mod} = 0$ region, all solutions (green line) are PT solutions, 
and a time-scaling law is observed as $\omega$ approaches $\omega^*$,  
\begin{equation}
T = \gamma \ln| \omega - \omega^*|,
\end{equation}
where $\gamma$ is the nonuniversal scaling exponent. 
Approaching the boundary of the $n_{\rm mod} = 0$ region from the other direction, one sees that 
regions of $n_{\rm mod} = 1$ PT solutions approach
the edge of the $n_{\rm mod} = 0$ region in a log-periodic fashion.  The time-minimum boundary
of each band follows a time-scaling law with the same exponent.
Between the $n_{\rm mod} = 1$ bands, one can also see a rich structure of BS solutions and additional 
PT bands with $n_{\rm mod} > 1$ (Figure \ref{fig:CritSolnZoom}). 
Both the $n_{\rm mod} = 1$ bands and the structure between bands can be seen 
in Figure \ref{fig:CritSolnOverlay} to repeat in a discretely self-similar fashion,
\begin{equation}
T(\ln|\omega-\omega^*|) = T(\ln|\omega-\omega^*| + n\Delta_\omega) + n\Delta_T,
\end{equation}
where $T(\ln|\omega-\omega^*| + n\Delta_\omega) + n\Delta_T$ has been plotted for 
$n=0,1,2,3,4,5,6,\rm{\ and \ } 7$.
While regular (i.e., not chaotic), this self-similarity in the 
 set of PT solutions is fractal in nature because it has a nonintegral Minkowski-Bouligand dimension
 that depends on the width of and spacing between the  $n_{\rm mod} = 1$ bands in $\ln|\omega-\omega^*|$  
 space, similar to \cite{Honda2010}.

\subsection{``Modest" Superextremal RN-AdS Solitons}

It was shown in Section \ref{sec:Coherent} that for the shape of potential ($\alpha_n$) 
and charge coupling ($q$) used in this paper, super-extremal solutions exist for 
$0.029 \lesssim \kappa_g \lesssim 0.035$. 
Given that solutions with $\kappa_g \gtrsim 0.0155$ were observed to be unstable, 
one might naturally wonder whether the long-term fate of such superextremal solutions
could be to collapse to form a naked singularity.

Because these solutions are bubbles with a true vacuum interior and a false vacuum exterior, 
local negative energy densities can be present inside these bubbles, and both the dominant
energy condition and the weak energy condition can be 
violated and cosmic censorship does not necessarily hold.
However, the only source of negative energy is from the scalar field potential, $V(\phi)$, and the 
amount of negative energy is proportional to the volume of space where $\phi \approx \phi_T$.
As such, if the bubble wall were to completely collapse, so would the volume of space that would contribute 
``negatively" to the local energy density.  
This would suggest that upon completely collapsing, the amount of negative energy arising from the true 
negative vacuum energy density would go to zero, the energy conditions could again be met, 
cosmic censorship would again hold, and naked singularities would not be present.

A more compelling argument can be made for the long-term ``modesty" (lack of nakedness) of these solutions
by considering the conservation of energy.
Every evolution studied in this paper begins with perturbed bound-state initial data.  The total matter, 
$M_\infty$, is of finite extent,  converges rapidly after $r\approx \xi$, and is a conserved quantity.
For the initial data with opposite-charged perturbations, the Coulombic self-repulsion is decreased 
and the bubble wall (and total mass) collapses inward.  
Unlike the subextremal case in which the bubble wall falls within the outer RN horizon, for this
superextremal case there are no RN horizons and the formation of a naked singularity depends
on whether the collapse continues indefinitely.
Because the collapsing mass distribution is {\it charged}, one must consider the energy it takes to 
compress that charge against its own Coulombic self-repulsion.  
Complete collapse of the charged matter to a naked singularity would require an infinite amount
of energy; 
this is analogous to the well-known infinite self-energy of a point charge.  
Because there is only a finite amount of energy, $M_\infty$, that can be converted to electromagnetic 
mass-energy, and since there is a finite (and {\it decreasing} with radius) amount of negative energy 
from the true vacuum,
the collapse must stop at some non-zero radius.
Appendix \ref{sec:AppendixA} describes simple models describing the possible end-state of 
such a collapse and derives four different minimum radii based on different assumptions about 
the shape of the charge distribution.
%

\begin{figure}[t]
\newcommand{\picxyB}{8cm}
\vspace{1mm}
\hbox{
\includegraphics[width=\picxyB,height=\picxyB]{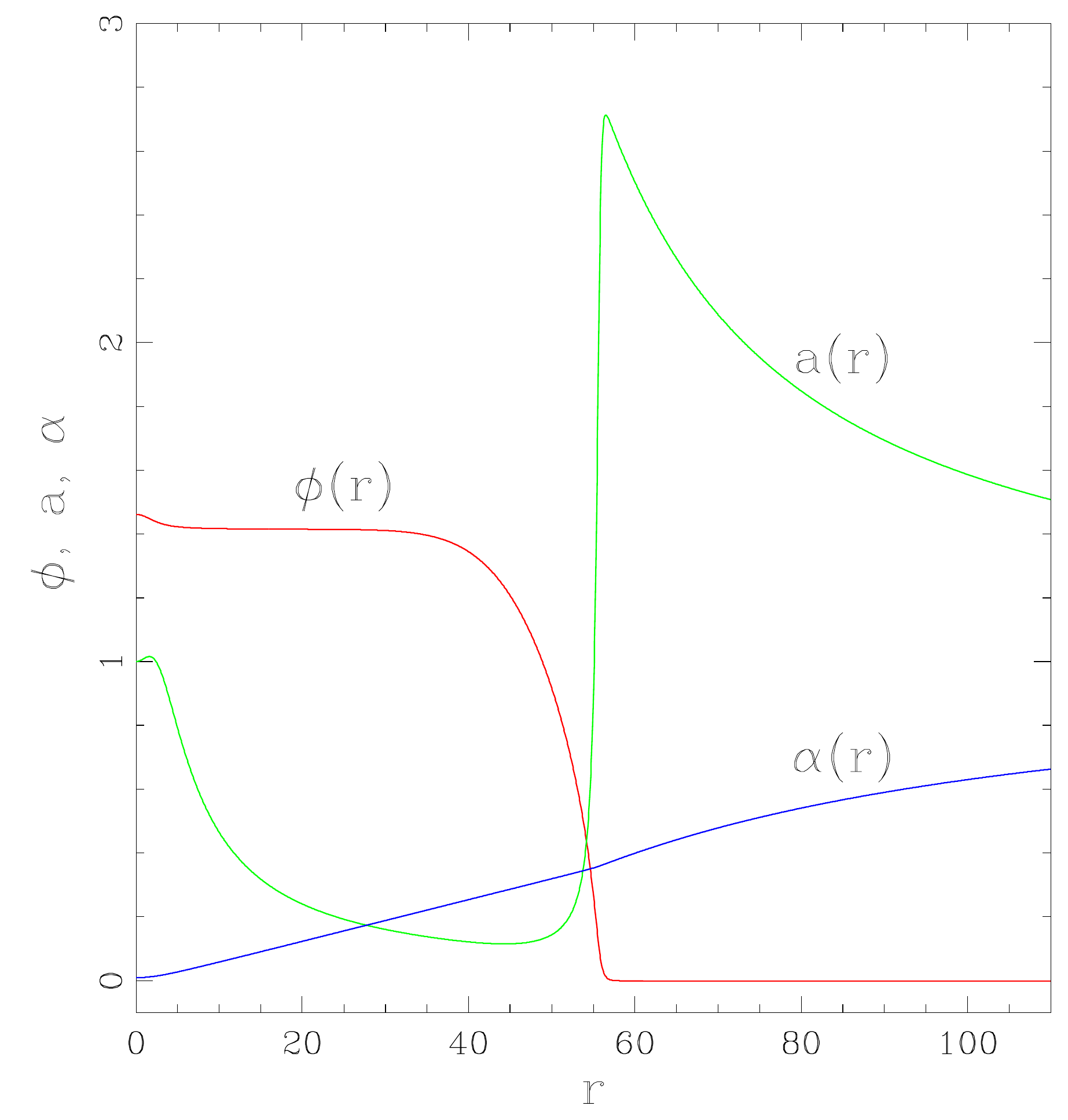}
}
\caption{  
Graphs of the scalar field, $\phi(r)$, geometry, $a(r)$, and lapse function, $\alpha(r)$, in red, green,
and blue, respectively, for $\kappa_g = 0.031$ and $u_0 = \pm 0.5$.  This solution is 
superextremal with $|\Xi| \approx 1.03$.
}
\label{fig:CoherentFields}
\end{figure}
\begin{figure}[t]
\newcommand{\picxyB}{8cm}
\vspace{1mm}
\hbox{
\includegraphics[width=\picxyB,height=\picxyB]{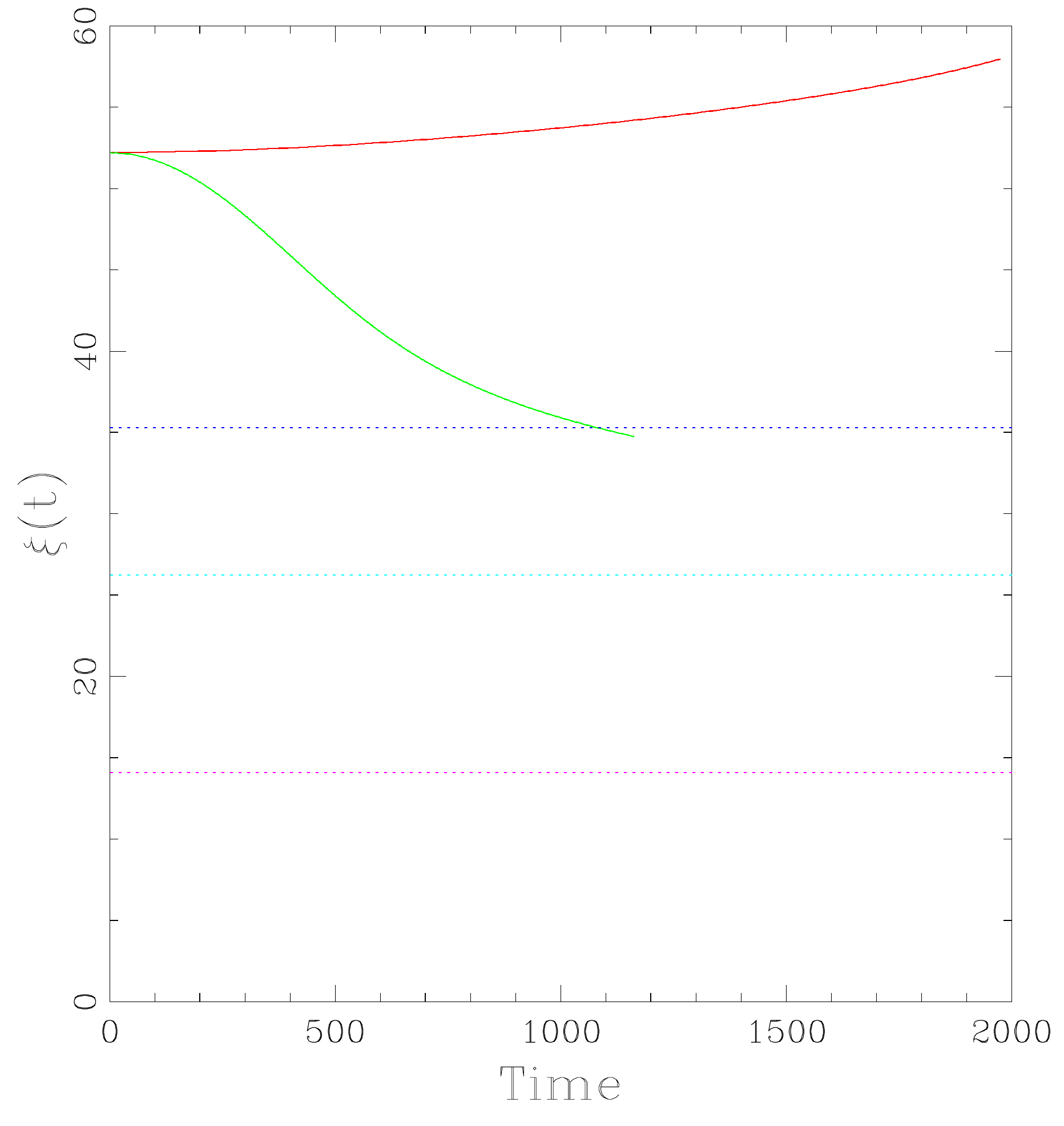}
}
\caption{  
Graphs of the bubble radius $\xi(t)$ as a function of time for a like-charged (red)
and opposite-charged (green) perturbation, for $\kappa_g = 0.031$ and $u_0 = \pm 0.5$.  
The like-charged perturbed solution expands, and the opposite-charged solution collapses.
The unperturbed solution is superextremal with $|\Xi| \approx 1.03$.  
The horizontal lines represent the minimum collapse radius based on conservation of energy for
a Gaussian $J^t$ (magenta), linear $J^t$ (cyan), constant $dQ/dr$ (blue), and constant $\rho$ (equal to 
constant $dQ/dr$).   
}
\label{fig:PerturbedFields}
\end{figure}
\begin{figure}[t]
\newcommand{\picxyB}{8cm}
\vspace{1mm}
\hbox{
\includegraphics[width=\picxyB]{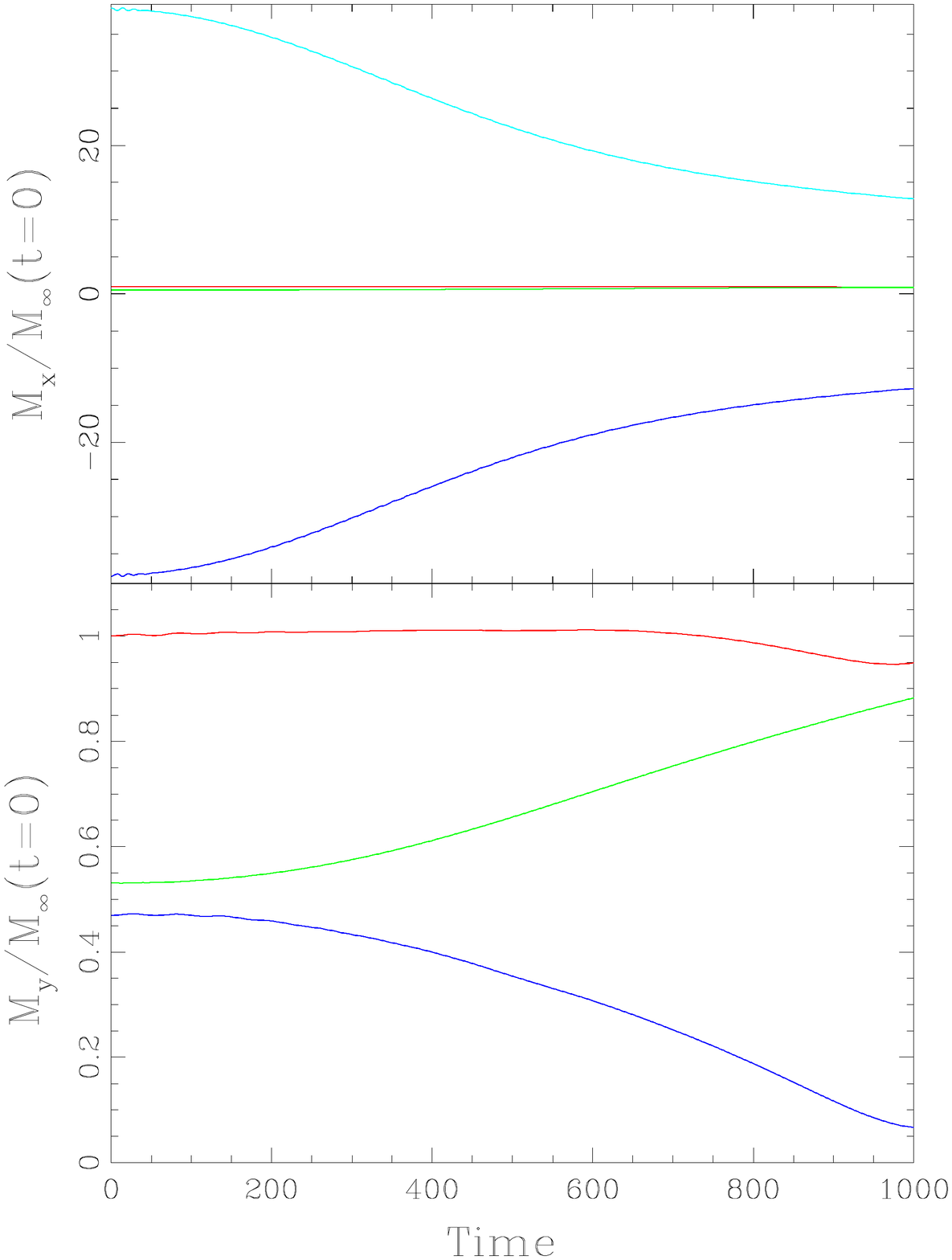}
}
\caption{  
Graphs of composite energies for charge-perturbed evolution of a superextremal RN-AdS
soliton with $\kappa_g = 0.031$ and $u_0 = \pm 0.5$.  The top graph shows 
$M_\infty$, $M_{\scriptscriptstyle\rm EM}$, $M_{\scriptscriptstyle\rm V}$, and
$M_{\scriptscriptstyle\rm D\phi}$ scaled by the initial total mass, in red, green, blue, and cyan, respectively.
The bottom graph shows $M_\infty$, $M_{\scriptscriptstyle\rm EM}$, and 
$M_{\scriptscriptstyle\rm V} + M_{\scriptscriptstyle\rm D\phi}$ scaled by the initial total mass
in red, green, and blue, respectively.
Total mass is conserved to within a percent through $t\approx 800$ and to within 
about 5 percent through $t\approx 1000$, but the solution becomes unstable shortly thereafter.
}
\label{fig:EnergyComparison}
\end{figure}

Figure \ref{fig:CoherentFields} shows the scalar field, geometry, and lapse function for a coherent solution with 
$\kappa_g = 0.031$ and $u_0 = \pm 0.5$, which has $|\Xi| \approx 1.03$.  
Within the bubble wall radius ($\xi \approx 52$), the scalar field is approximately $\phi_T$, 
the geometric variable $a(r)$ can be seen to decrease from its value of unity 
at the origin (required by elementary flatness and regularity), and the lapse function
can be seen to increase roughly linearly.  
As the solutions approach the bubble wall, the fields transition from their AdS values 
and match onto an RN exterior.
Figure \ref{fig:PerturbedFields} shows the dynamics resulting from like-charged and 
opposite-charged perturbations of this solution.  
As expected, like-charged solutions form expanding bubbles that induce a phase transition, 
while opposite-charged solutions collapse.

To better understand the dynamics of collapsing superextremal solutions, it is helpful 
to understand the composite masses that contribute to the total ADM mass,
$M_\infty$. 
One can define 
\begin{eqnarray}
M_{\scriptscriptstyle\rm V}	&=&   4\pi \int_0^{r_b} dr r^2 V ,\\
M_{\scriptscriptstyle\rm EM}	&=&   4\pi \int_0^{r_b} dr r^2 \left( \frac{E_r^2}{2}\right),  {\rm and}\\
M_{\scriptscriptstyle\rm D\phi}	&=&   M_\infty - M_{\scriptscriptstyle\rm EM}	-M_{\scriptscriptstyle\rm V},
\end{eqnarray}
for the scalar potential, the electromagnetic field, and the gauge covariant derivative of the scalar field, 
respectively.
One can see in Figure \ref{fig:EnergyComparison} that for this particular  RN-AdS bubble solution,
the magnitudes of $M_{\scriptscriptstyle\rm V}$ and $M_{\scriptscriptstyle\rm D\phi}$
are many times larger than the magnitude of the total ADM mass, 
$M_\infty$.
The vacuum energy inside the bubble provides a very large negative energy contribution, 
approximately $-38.03 M_\infty$, while
the gauge covariant derivative energy provides a very large (slightly larger in magnitude) positive 
energy contribution,
approximately $38.50 M_\infty$.  The sum of these two terms is approximately $0.47 M_\infty$, and the
remaining mass, approximately $0.53 M_\infty$, is mass-energy from the electromagnetic field.
As the wall begins to collapse, $M_{\scriptscriptstyle\rm EM}$ begins to increase toward $M_\infty$, 
and  $M_{\scriptscriptstyle\rm V}+M_{\scriptscriptstyle\rm D\phi}$ begins to decrease.
Scalar field energy is being converted into electromagnetic energy because the fields are doing work
by compressing the charge against Coulombic repulsion.
Unfortunately, solutions become unstable around $t\approx 1150$ as a result of {unphysical}
gauge shocks that arise from gauge radiation ($A_t$ and $A_r$). 
The infalling radiative parts of the gauge potential do not contribute to $E_r$ but become severely 
compressed by the nearly collapsed lapse ($\alpha_0 \lesssim 10^{-5}$), leading to steep gradients and 
code instability.
Nevertheless, there is clear evidence that superextremal solutions with opposite-charged 
perturbations can collapse and result in strong-field gravitational solutions that 
match onto superextremal RN exteriors with  nonsingular interiors.  
%


\section{Conclusions}

Results have been presented from numerical simulations of
the EMH equations with a broken U(1) symmetry.
Coherent nontopological solitons were shown to exist that separate a negative-energy 
AdS true vacuum  interior ($V(\phi_T)<0$) from a zero-energy RN 
false vacuum exterior  ($V(\phi_F)=0$).
The physical parameters (charge, mass, radius, and central lapse) of these solutions were obtained
for a wide range of gravitational couplings.
For $\kappa_g \gtrsim 0.011$, solutions are gravitationally strong-field solutions with radii on the order of their 
outer RN horizons and with central lapse function values of $0.1\lesssim\alpha_0\lesssim 0.5$, indicating significant 
gravitational time dilation effects.
For solutions with $0.028 \lesssim \kappa_g \lesssim 0.034$, the charge-to-mass ratios become
superextremal ($|\Xi| >1$), the radii are $(\kappa_g M_\infty) \lesssim \xi  \lesssim (3\kappa_g M_\infty)$, 
and the central lapse values are $0.001\lesssim\alpha_0\lesssim 0.08$.
Because obtaining these solutions requires fine-tuning to more than one part in $10^{15}$, 
a 96-bit precision numerical code was used.  

The stability of these solutions was tested by perturbing the charge of the coherent solution and evolving
the time-dependent equations of motion.
In the weak gravitational limit, the short-term stability depends on the sign 
of $(\omega/ Q) \, \partial_\omega Q$, similar to Q-balls.
This condition does not hold, however, for  $\kappa_g \gtrsim 0.015$ and for 
$|u_0| \gtrsim 0.6$. 
The long-term end-states of the perturbed solutions were visualized using  ``phase diagrams"
that served as a way to clearly demonstrate regions of stability and instability.
It was further demonstrated that there exists a rich fractal structure around the ``triple point,"
which was defined to be a region in the phase diagrams where BH, PT, and BS 
solutions exist in close proximity.
The fractal structure results from bounce-like modulations of the scalar field where collapsing 
bubbles  bounce back and induce phase transitions to the AdS true vacuum.  
The bands of $n$ modulations are surrounded by bands of $n+1$ modulations that approach the 
boundary of the $n$ modulation regions in a log-periodic manner.  
Threshold solutions are shown to demonstrate time-scaling laws with scaling exponents that depend 
on $\kappa_g$ and total charge.

Finally, superextremal coherent RN-AdS solitons were shown to be unstable
with an end state dependent on the sign of the charge perturbation.
Like-charged perturbations led to expanding bubbles, 
and opposite-charged perturbations led to collapsing bubbles.  
While the superextremal charge-to-mass ratio might suggest a possibility of collapse to a naked singularity,
it was shown that there is a minimum radius within which the wall can collapse based on the conversion 
of scalar field energy to electromagnetic energy.
While it is still possible that solutions collapse and then bounce back to induce a phase transition,
there is strong evidence supporting the existence of persistent nonsingular superextremal
bound states that exhibit very strong-field gravitational behavior with
$\alpha_0 \lesssim 10^{-5}$.
%
%
%
The existence of such solutions and their formation by decay of unstable coherent RN-AdS solitons 
was not previously known.

\begin{appendix}

\section{Units and Dimensions}
\label{sec:AppendixB}

The action described in equation (\ref{eqn:EMHAction}) 
was derived from the following dimensionful Lagrangian 
where $c=\hbar=\epsilon_0=1$:
\begin{eqnarray}
L&=&\frac{R}{16\pi G_N} -\frac{F_{\mu\nu}F^{\mu\nu}}{4}
-\frac{1}{2}g^{\mu\nu}\left( D_\nu\phi\right)^*\! D_\mu\phi \nonumber\\
&&-  \frac{1}{2}\alpha_1 m^2\phi^2 
-  \frac{1}{4}\alpha_2 \phi^4 
-  \frac{1}{6}\alpha_3 m^{-2}\phi^6,
  \end{eqnarray}
where $m$ is the characteristic boson mass, $G_N$ is Newton's constant, 
and the $\alpha_n$ are dimensionless.

%
Equation (\ref{eqn:EMHAction}) is derived by transforming all the field variables and coordinates 
to dimensionless quantities according to the following transformations:
\begin{eqnarray}
\tilde{r}	&=& m r\\
\tilde{t}	&=& m t\\
\tilde{A}_\mu	&=& m^{-1} A_\mu \\
\tilde{\phi}	&=& m^{-1} \phi
\end{eqnarray}
which puts the Lagrangian in the following dimensionless form:
\begin{eqnarray}
\frac{1}{m^4}L&=& \frac{\tilde{R}}{16\pi \kappa_g} -\frac{\tilde{F}_{\mu\nu}\tilde{F}^{\mu\nu}}{4}
-\frac{1}{2}g^{\mu\nu}\left( \tilde{D}_\nu\tilde{\phi}\right)^*\! \tilde{D}_\mu\tilde{\phi} \nonumber\\
&&-  \frac{1}{2}{\alpha}_1 \tilde{\phi}^2 
-  \frac{1}{4}{\alpha}_2 \tilde{\phi}^4 
-  \frac{1}{6}{\alpha}_3\tilde{\phi}^6 \label{eqn:tildeL} \\
&=& \tilde{L} 
\end{eqnarray}
where $\kappa_g = m^2 G_N$ is dimensionless.
Removing tildes in equation (\ref{eqn:tildeL}) results in equation (\ref{eqn:EMHAction}).

\section{Minimum Radius of Gaussian Charge Distribution}
\label{sec:AppendixA}

This appendix discusses a simplified model of a collapsing charge distribution and uses the 
conservation of energy to derive the minimum radius to which the distribution can collapse.
%
The charge density  is assumed to be
\begin{eqnarray}
J^t &=& 	\frac{Q_0}{\sigma^3 \pi^{3/2}}{\rm e}^{-\frac{r^2}{\sigma^2}},
\end{eqnarray}
which is normalized so that the total charge (enclosed) at infinity is $Q_0$.   
The charge enclosed for arbitrary $r$ is given by
\begin{eqnarray}
Q(r_0) 	&=& 	4\pi \int_0^{r_0} dr r^2 J^t\\
		&=& Q_0 \left[   
		-\frac{2 r_0 }{\sigma \pi^{1/2}}{\rm e}^{-\frac{r_0^2}{\sigma^2}} + {\rm erf}\left(\frac{r_0}{\sigma}\right)
		\right].
\end{eqnarray}
Integrating equation (\ref{eqn:CoherentEr})  from the origin to $r_0$ gives
\begin{eqnarray}
E_r(r_0) 	&=&  \frac{Q(r_0)}{4 \pi r_0^2} \\
		&=& 	\frac{Q_0}{4 \pi  r_0^2}  
		\left[   
		-\frac{2 r_0 }{\sigma \pi^{1/2}}{\rm e}^{-\frac{r_0^2}{\sigma^2}} + {\rm erf}\left(\frac{r_0}{\sigma}\right)
		\right].
\end{eqnarray}
The mass-energy of the electromagnetic field is then given by
\begin{eqnarray}
M_{\scriptscriptstyle\rm EM}(r_0)\!\! 
		&=& 	\!\! 	4\pi \int_0^{r_0} dr r^2 \left( \frac{ E_r(r_0)^2}{2}\right) \\
		&=& 	\!\!
		-\frac{Q_0^2}{8 \pi  r_0}  
		\left[   
			{\rm erf}\left(\frac{r_0}{\sigma}\right)^2 \!\!
			- \sqrt{\frac{2}{\pi}}\frac{r_0}{\sigma} {\rm erf}\left(\frac{\sqrt{2}r_0}{\sigma}\right)
		\right].  \label{eqn:GeneralMassofDeltaCharge} \nonumber \\
		&& 
\end{eqnarray}
%
Integrating the Hamiltonian constraint equation (\ref{eqn:CoherentA}) yields the well-known equation 
for the geometry
\begin{eqnarray}
a^2	&=& \left( 1 - \frac{2 \kappa_g M(r)}{r}\right)^{-1} \\
	&\equiv& \Delta^{-1},
\end{eqnarray}
where $M(r)$ is the total ADM mass,
\begin{eqnarray}
M(r) = M_{\scriptscriptstyle\rm D\phi}(r) +M_{\scriptscriptstyle\rm V}(r) + M_{\scriptscriptstyle\rm EM}(r),
\end{eqnarray}
and where
$M_{\scriptscriptstyle\rm D\phi}(r)$ is the mass-energy term arising from  the covariant derivative of the scalar field,
and $M_{\scriptscriptstyle\rm V}(r)$ is the mass-energy arising from the scalar field potential. 
%
For  $r_0 \gg \sigma$, the mass-energy of the electromagnetic field becomes
\begin{eqnarray}
M_{\scriptscriptstyle\rm EM}(r_0 \gg \sigma) 	
		&=& 	
		-\frac{Q_0^2}{8 \pi r_0}  
		+ \frac{Q_0^2}{4\sqrt{2} \pi^{3/2}  \sigma},    \label{eqn:MEM_rggdelta}
\end{eqnarray}
where the first term is the typical  RN charge term, 
and the second term is a  \emph{positive} constant that can be considered the mass of the electromagnetic field
at infinity, $M^\infty_{\scriptscriptstyle\rm EM}$:
\begin{eqnarray}
\Delta(r_0 \gg \sigma) 	&=& 1- \frac{2\kappa_g M^\infty_{\scriptscriptstyle\rm EM}}{r_0}  
+ \frac{\kappa_gQ_0^2}{4\pi r_0^2}, \label{eqn:myRN}
\end{eqnarray}
where 
\begin{eqnarray}
M^\infty_{\scriptscriptstyle\rm EM}	&=&  \frac{Q_0^2}{4\sqrt{2} \pi^{3/2}  \sigma}.
\end{eqnarray}
It should be stressed that the $r_0 \lesssim \sigma$ behavior of equations
(\ref{eqn:MEM_rggdelta}) and (\ref{eqn:myRN}) is not valid because  the assumption
used to obtain the expression is violated in that regime.  
The  $r_0$-dependent terms are only proportional to $r_0^{-1}$ and $r_0^{-2}$  
at large $r_0$.  The actual expression for the mass (\ref{eqn:GeneralMassofDeltaCharge}) is well 
behaved (approaches zero) as $r_0$ goes to zero.  
%
%

The total mass of the field, however, does indeed go to infinity in the limit 
of small $\sigma$.  
This is consistent with the well-known infinite self-energy of a point charge.
Nevertheless, the conservation of energy would imply that the final mass-energy of the electric field 
cannot exceed the total conserved ADM mass of the system:
\begin{eqnarray}
M^{\scriptscriptstyle\rm final}_{\scriptscriptstyle\rm EM} &\leq& M_{\scriptscriptstyle\rm ADM},
\end{eqnarray}
which implies
\begin{eqnarray}
\sigma  &\geq& 
\frac{1}{4\sqrt{2} \pi^{3/2}  }  
\left(\frac{Q_0^2}{ M_{\scriptscriptstyle\rm ADM}   }  \right), 
\end{eqnarray}
which would set a limit on the collapse of the charge distribution.  

While the minimum radius for a Gaussian-shaped charge density with mass $M_{\scriptscriptstyle\rm ADM}$ 
is derived above, 
Table \ref{table:MinCollapseRadii} presents the results of three additional models that all result 
in comparable minimum radii.

\begin{table}[t]
\begin{center}
\vspace{10mm}
\begin{tabular}{lcr}
\hline
\hline
Model Description & \hspace{12mm} &  Constant ($k$) \\
\hline
Gaussian $J^t$ & & $\displaystyle 1/ (4\sqrt{2}\pi^{3/2} )$ \\
Linear $J^t$ & & $\displaystyle 13/(70 \pi)$ \\
Constant $dQ/dr$ & & $\displaystyle  1/(4\pi)$ \\
Constant $\rho \approx V(\phi_{\rm T})$, w/ approx. & & $\displaystyle  1/(4\pi)$ \\
\hline
\end{tabular}
\caption{Table of constants to determine a lower limit on the  radius for collapsed
RN-AdS solutions, where the constant $k$ is defined by 
$\sigma_{\rm min} = k Q_0^2/M_{\scriptscriptstyle\rm ADM}$.
}
\label{table:MinCollapseRadii}
\end{center}
\end{table}

\end{appendix}


\end{document}